\documentclass{statsoc}
\usepackage{amssymb,amsmath}
\usepackage{graphicx,bbm,pstricks}
\usepackage[left=1in,right=1in,top=1in,bottom=1in]{geometry}
\usepackage{bbm,multirow}
\usepackage{algorithmic}
\usepackage{float}
\usepackage{xcolor}
\usepackage[colorinlistoftodos]{todonotes}
\usepackage{tikz}
\usepackage{hyperref}
\usetikzlibrary{shapes,arrows}
\newfloat{Algorithm}{t}{lop}
\floatstyle{boxed}
\restylefloat{Algorithm}

\newcommand{\edit}[1]{{#1}}
\newcommand{\editfinal}[1]{{#1}}

\newcommand{\E}{\operatorname{E}}
\newcommand{\Var}{\operatorname{Var}}

\newtheorem{theorem}{Theorem}
\newtheorem{lemma}{Lemma}

\title[Efficient forecasting and UQ for debt recovery simulations]{Efficient forecasting and uncertainty quantification for large scale
  account level Monte Carlo models of debt recovery}
  \author{Sam Baynes}
  \address{Arrow Global Ltd, Manchester, UK}
  \email{sbaynes@arrowglobal.net}
  \author{Simon L. Cotter}
  \address{University of Manchester, Manchester, UK}
  \email{simon.cotter@manchester.ac.uk}
  \author{Paul Russell}
  \address{Arrow Global Ltd, Manchester, UK}
  \email{prussell@arrowglobal.net}
  \author{Edmund Ryan}
  \address{University of Manchester and Arrow Global Ltd, Manchester, UK}
  \email{edmund.ryan@corndel.com}
  \author[Sam Baynes {\it et al}]{Tim Waite}
  \address{University of Manchester, Manchester, UK}
  \email{timothy.waite@manchester.ac.uk}

\begin{document}
  
\begin{abstract} 
\edit{We consider the problem of forecasting debt recovery from large portfolios of non-performing unsecured consumer loans under management. The state-of-the-art in industry is to use stochastic processes to approximately model payment behaviour of individual customers based on several covariates, including credit scores and payment history. Monte Carlo simulation of these stochastic processes can enable forecasting of the possible collections from portfolios of defaulted debt.

Even though the individual-level models are relatively simple, it is challenging to carry out simulations at the portfolio level because of the very large number of heterogeneous accounts, with a broad range of values for the collection variances.

We aim to solve two main problems: efficient allocation of computational resources in the simulations to estimate the likely collections as precisely as possible, and quantification of the uncertainty in the forecasts, under the constraint that all the accounts must be simulated to enable valuation at the account level. We show that, under certain conditions, robust estimators of population-level variance can be constructed by summing over coarse unbiased estimators of the variance of individual accounts. The proposed methods are demonstrated through application to a model which shares key features with those that are used in practice.}

\end{abstract}
%65C05 Monte Carlo methods
%62P05 Applications of statistics to actuarial sciences and financial mathematics
%
  
\section{Introduction}
%
% What about the following updated intro?
%

Stochastic models are an important tool in the financial industry for understanding both expected outcomes and the uncertainty or risk associated with these outcomes. Though the academic literature on models for traditional areas such as stocks, insurance, and pensions is well-developed \cite{micciche2002volatility,beard2013risk,battocchio2004optimal}, other applications such as non-performing loans (NPLs) \edit{(any loan in which the original terms of the loan have been breached, e.g. missed or lower payments than in the agreement)} are less widely studied. \editfinal{A few important examples of NPL modelling exist, e.g. \cite{thomas2016modelling}, where Markov models are developed for repayment sequences beginning with a default, and \cite{bellotti2021forecasting} where machine learning methods are used, however overall the literature is less well-developed. Despite this, NPLs are a highly commercially important application.}

Arrow Global Ltd (AGL) is a leading European investor and asset manager, with millions of non-performing unsecured consumer loans under management, and is the leading purchaser of non-performing consumer debt portfolios in the UK. Stochastic modelling of the remaining future collections from both existing and prospective portfolios is critical for the company’s financial reporting, purchasing decisions, and its decisions about operational strategies, for example under what circumstances to pursue litigation. 

The key feature of Arrow’s problem is the need to aggregate detailed models of individual-level outcomes to make forecasts at the population or portfolio level \edit{on a biannual basis, each of which is termed a \emph{forecasting round}.} This introduces several challenges. Monte Carlo simulation is required because the quantities of interest such as the expectations and quantiles of the collections are analytically intractable. However, the simulation of portfolio-level outcomes is a highly computationally expensive process due to the size of a typical portfolio, which may consist of tens or hundreds of thousands of heterogeneous accounts. This large scale forces a high cost for portfolio simulation even though Arrow’s individual-level model is relatively simple, being essentially a Markov model similar to \cite{thomas2016modelling} but using a logistic regression to model an individual’s probability of payment in a given period as a function of the account attributes. As a result of this high expense the number of simulation realisations is highly constrained, and it is important to make efficient use of the available computational resources. In addition, due to the restricted number of realisations available it is not clear at present how to use the simulations to quantify the uncertainty in the portfolio-level forecasts. \edit{Moreover, we live under the constraint that all accounts must be simulated within the forecast, in order to allow valuations at the account level. This constraint rules out methods based on sub-sampling, which could be employed to deal with the large size of the population. Furthermore, due to the heterogeneity of the accounts, it is important that certain sub-populations have sufficient numbers of realisations of simulations.}

In this paper we address these challenges with three main methodological developments. First, we devise more efficient Monte Carlo schemes that reduce the variance of the estimates by optimizing the number of realisations of the different accounts in the simulation. The method parallels similar ideas from Multilevel Monte Carlo \cite{giles2008multilevel,giles2015multilevel} and Bayesian design of experiments \cite{beck2018fast}.  The optimal number of realisations is shown to depend on the variance of the account-level outcome, which is a function of the account attributes. Second, we develop prediction intervals to allow uncertainty quantification in the aggregated forecasts. Third, in order to facilitate a practical implementation of both the prediction intervals and the optimal Monte Carlo scheme a Gaussian process emulator \cite{sacks1989design,gramacy2020surrogates} is developed to give a computationally cheap estimate of the variance of the account-level collection as a function of the account attributes, such as credit score, operational segment, etc. 

Our focus is not on the construction of realistic individual-level models, which can be accomplished using fairly standard techniques. Rather, the primary aim of the paper is to develop variance reduction and uncertainty quantification techniques in the case where individual-level forecasts are aggregated to produce population-level forecasts. This structure appears fairly generic and so we expect our techniques to be applicable beyond the context of Arrow and consumer debt. Due to its commercially sensitive nature the full details of Arrow's commercial model cannot be divulged and so we instead illustrate the application of our variance reduction and uncertainty quantification techniques with a representative agent-based model that reflects the main features of the commercial model.  

The structure of the paper is as follows. In Section \ref{sec:TM} we introduce the representative model and the existing method for producing population-level forecasts from individual simulations. In Section \ref{sec:Err} we discuss techniques for variance reduction and illustrate their effectiveness on the representative model. In Section \ref{sec:UQ} we present the prediction intervals, their asymptotic properties, and simulation studies validating their coverage. In Section \ref{sec:protect} we introduce methodology for protecting quality of the forecasts in important subsets of the full population. In Section \ref{sec:GP-predict-variance} we give details of our Gaussian process emulation methodology.  We finish with conclusions and discussion in Section \ref{sec:dicuss}.

\section{Representative Model}\label{sec:TM}

AGL has developed a suite of expected recovery forecasting models for their secured and unsecured European assets. These forecasting models typically predict recovery curves of at least seven years for these assets. In this paper, we consider specifically the UK unsecured consumer debt model. \edit{This model simulates monthly cash payments for each account contained within the population,} using knowledge of each account's credit rating, past engagement behaviour, and the efficacy of AGL's collection strategies. The account population has accumulated over more than a decade and is composed of a wide variety of products, from loans, overdrafts, and credit cards to mail order credit, telecommunication debt, and student loans. As such the portfolio is highly heterogeneous, and the model is  complex.

The methods described in this paper were designed and implemented using AGL's commercial forecasting model. However, in order to both simplify the discussion and protect AGL's intellectual property here we will illustrate their use on a representative model that shares the main characteristics of AGL's model.

\subsection{Model overview}

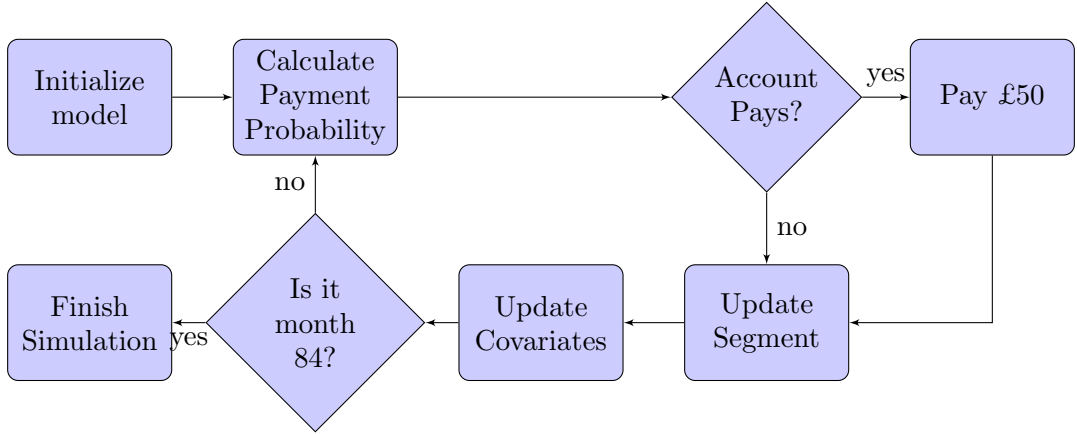
\begin{figure}
\centering
\begin{tikzpicture}[node distance = 3cm
      , auto
      , decision/.style = {diamond, draw, fill=blue!20, 
    text width=4.5em, text badly centered, node distance=3cm, inner sep=0pt}
      , block/.style = {rectangle, draw, fill=blue!20, 
    text width=5em, text centered, rounded corners, minimum height=4em}
      , line/.style = {draw, -latex'}
                   ]

    \node [block] (init) {Initialize model};
    \node [block, right of=init] (PPay) {Calculate Payment Probability};
    \node [block, below of=init] (end) {Finish Simulation};
    \node [decision, right of=end] (CheckEnd) {Is it month 84?};
    \node [block, right of=CheckEnd] (update) {Update Covariates};
    \node [block, right of=update] (OpSeg) {Update Segment};
    \node [decision, above of= OpSeg] (MCDecision) {Account Pays?};
    \node [block, right of= MCDecision] (Pay) {Pay £50};

    \path [line] (init) -- (PPay);
    \path [line] (PPay) -- (MCDecision);
    \path [line] (MCDecision) -- node {yes}(Pay);
    \path [line] (Pay) |- (OpSeg);
    \path [line] (MCDecision) -- node {no} (OpSeg);
    \path [line] (OpSeg) -- (update);
    \path [line] (update) -- (CheckEnd);
    \path [line] (CheckEnd) -- node {yes}(end);
    \path [line] (CheckEnd) -- node {no}(PPay);
\end{tikzpicture}
\caption{A summary of the workflow of the representative model.}% \slc{This figure (and caption) needs updating by Paul, in particular to include possibility of segment transition. We may want to put this figure after 2.2.1 if we do this?}}
\label{fig:model-summary}
\end{figure}

The model we discuss here takes an ensemble of $N \in \mathbb{N}$ accounts, and evolves their characteristics (payment amount, balance, paying status) through a seven-year period, month by month. Therefore, one simulation of the model will produce $N$ time series, each containing up to 84 time points. From this collection of time series, we calculate the quantities of interest (QoI), which are the aggregate monthly collections and the total seven-year collection figure. Figure \ref{fig:model-summary} gives a visual summary of representative model, while full details are given in Section \ref{sec:detailed-model-description}.

At each time step the behaviour (i.e. payment or non-payment) of each individual account is simulated from a probability model that takes into account the account's credit score as well as their past engagement. 
In addition to these account characteristics, the payment probability will also depend on the collections strategy being applied by AGL to the account, e.g. an account may be worked amicably by agents in the call centre, be being prepared for enforcement action, or be undergoing enforcement actions through the courts. 
Within the model, these collections strategies are referred to as \emph{segments}.  

In AGL's model accounts can transition between segments, representing a decision to change between predefined strategies. The decisions about which accounts to transition is limited by the operational capacity, forcing an account comparison which leads to a source of dependence between the accounts. Accounts, once eligibility has been determined, are prioritised based on maximising the return on investment. This is an important feature and so we include segment transition dynamics within the representative model, detailed in Section \ref{sec:transitions}. 

%\slc{This needs editing once the interdependence of accounts in segment ? has been finalised,}

\subsection{Detailed model description}
\label{sec:detailed-model-description}
Some notation is required before the model can be clearly stated. Let $X_{it}$ denote the collections from the $i$th account in month $t$ ($i=1,\ldots,N$; $t=1,\ldots,84$). Further let $Y_{it} = I(X_{it}>0)$ denote the corresponding payment indicator and \edit{$\mathbf{Z}_{it}=(B_{it}, C_{i}, S_{it}, E_{i})$ denote a vector of covariates, with $B_{it}$ being the individual's remaining balance, $C_{i}$ their credit score, $S_{it}$ the operational segment in which the account resides, and $E_{i}$ the indicator for eligibility for segment transitions (see next section). Note that $C_i$ and $E_i$ do not change in time, unlike the other variables.} In order to initialize the model we will require knowledge of whether the account paid in the month preceding the simulation period; the corresponding indicator variable is denoted $Y_{i0}$.

At the individual level the main component of the representative model is essentially a logistic regression time series model. Namely it is assumed that given the covariates the conditional probability of a payment, 
$
p_{it} = P(Y_{it} = 1| Y_{i(t-1)}, \ldots, Y_{i0}, \mathbf{Z}_{it})
$,  
satisfies \edit{
$$
\log \frac{p_{it}}{1-p_{it}} = 
\begin{cases}
  -1 + 0.1 C_{i} + 2 Y_{i\,(t-1)} \,, &     \text{ if } S_{it} =1 \text{ and } B_{it} >0 \,, \\[0.5ex]
  0.4 C_{i} + 2 Y_{i\, (t-1)} \,, &         \text{ if } S_{it} =2 \text{ and } B_{it} >0\,, \\[0.5ex]
  -4 + 0.2 C_{i} + 2Y_{i\, (t-1)} \,, &     \text{ if } S_{it} =3 \text{ and } B_{it} >0 \,.
\end{cases}
$$}
\edit{In AGLs models, the different segments correspond with different collection strategies, and are another key source of the heterogeniety of account performance. Whilst the segments described above do not have real world meaning, they replicate the heterogeniety that AGLs model segments generate.}
While in the real-world model, payment amounts are extensively modelled, in the representative model if the customer pays, it is assumed that the payment amount is the lower of \pounds50 and the outstanding balance, hence overall $X_{it} = \min(50, B_{it}) Y_{it}$. The balance is updated dynamically according to $B_{i(t+1)} = B_{it} - X_{it}$. If the customer has no outstanding balance then there is no payment, i.e. if $B_{it}=0$ then $Y_{it} = X_{it} = 0$.

In the representative model the covariate values for $N$ accounts are initialized via simulation from appropriate distributions to give heterogeneous behaviour similar to that observed in AGL's commerical model.  Specifically, initial values for the payment indicator and balance are generated according to $Y_{i0} \sim \operatorname{Bernoulli}(0.2)$ and  $B_{i1}  \sim N(2500, 1000^2)$, the latter truncated to the range $[500,10000]$. The initial segments are chosen according to
$$
S_{i1} = \begin{cases}
       1 \,,& \text{ with probability } 0.2\,,\\
       2 \,,& \text{ with probability } 0.2\,,\\
       3 \,,& \text{ with probability } 0.6\,.
      \end{cases}
$$
\edit{The credit scores are initialized via sampling from a normal mixture $C_{i} \sim 0.15 \,N(1,1) + 0.05 \,N(4,1) + 0.2\, N(-1,1) + 0.6\, N(-5, \sqrt{0.1})$ and held fixed throughout the simulation.}

\edit{The use of a mixture distribution for credit score reflects the reality that Arrow's credit metrics exhibit clustering defined by key variables such as proportion of credit lines in good order, mortgage line in good order, any unsettled CCJs, etc., with less influential attributes creating variation within these clusters. The specific choice of a normal mixture is for convenience. Note that Arrow use custom credit metrics derived from the information in customer credit files, they do not use an off-the-shelf credit score product. }

\begin{figure}
\centering
\includegraphics[width=0.8\linewidth]{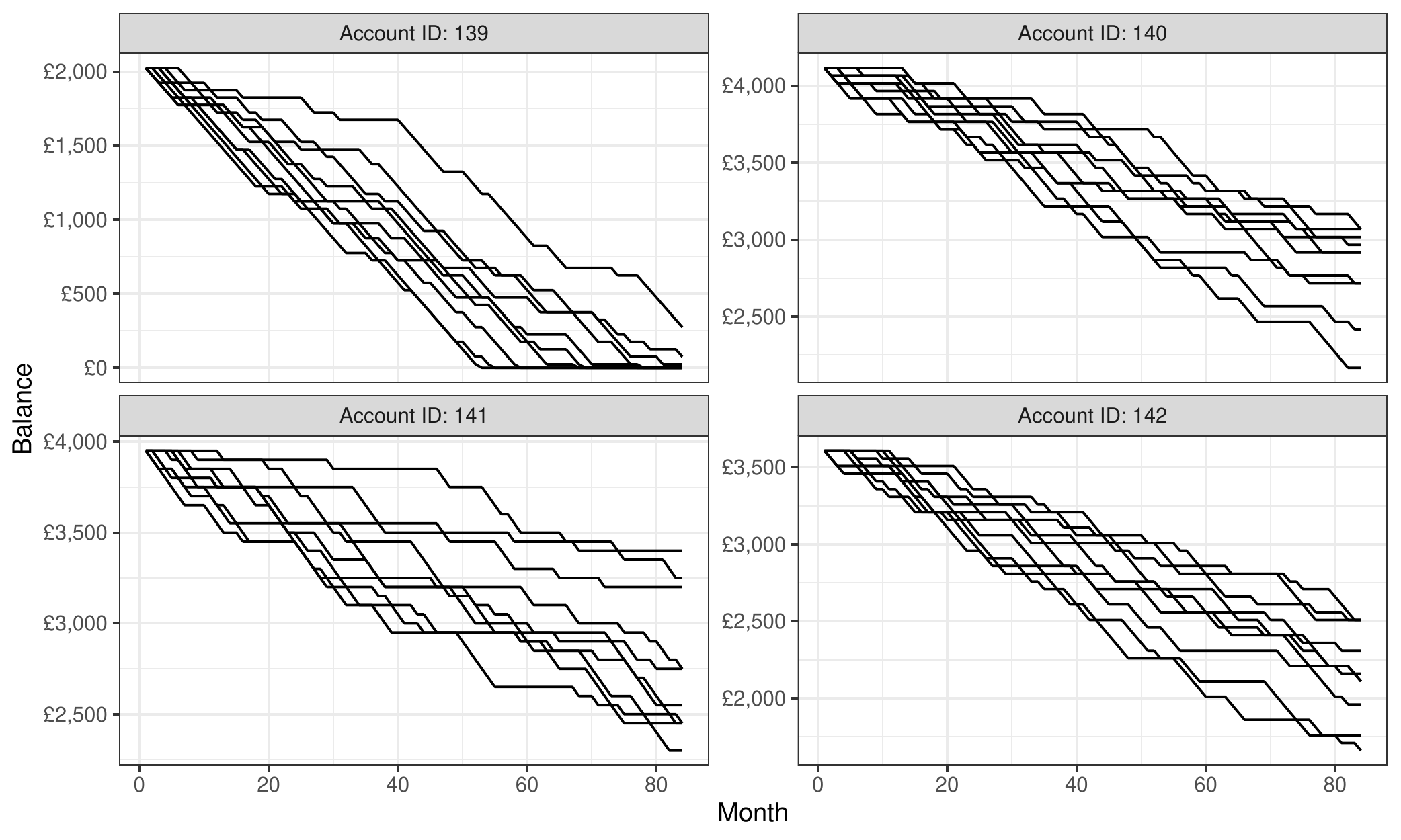}\\
\includegraphics[width=0.8\linewidth]{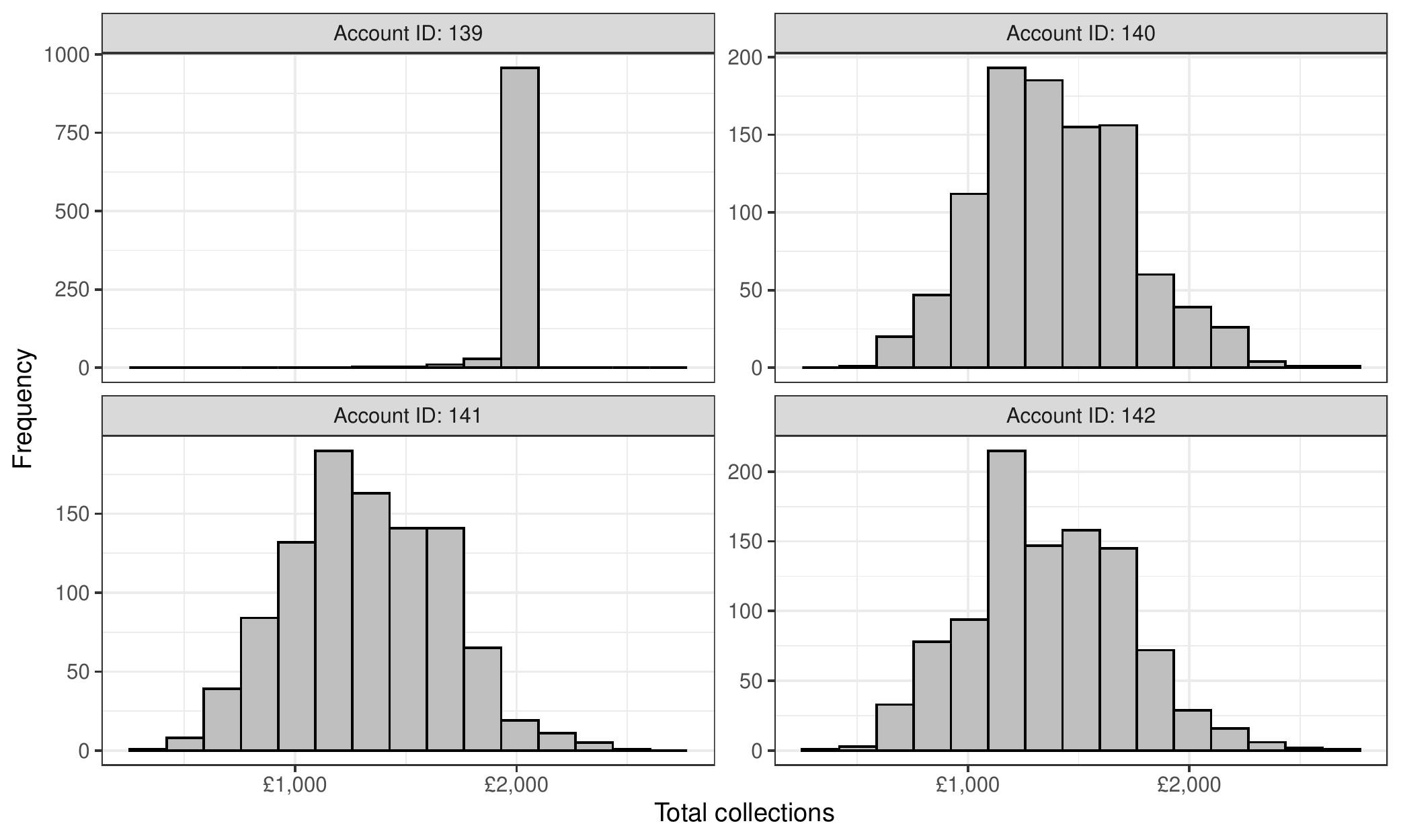}\\
\includegraphics[width=0.8\linewidth]{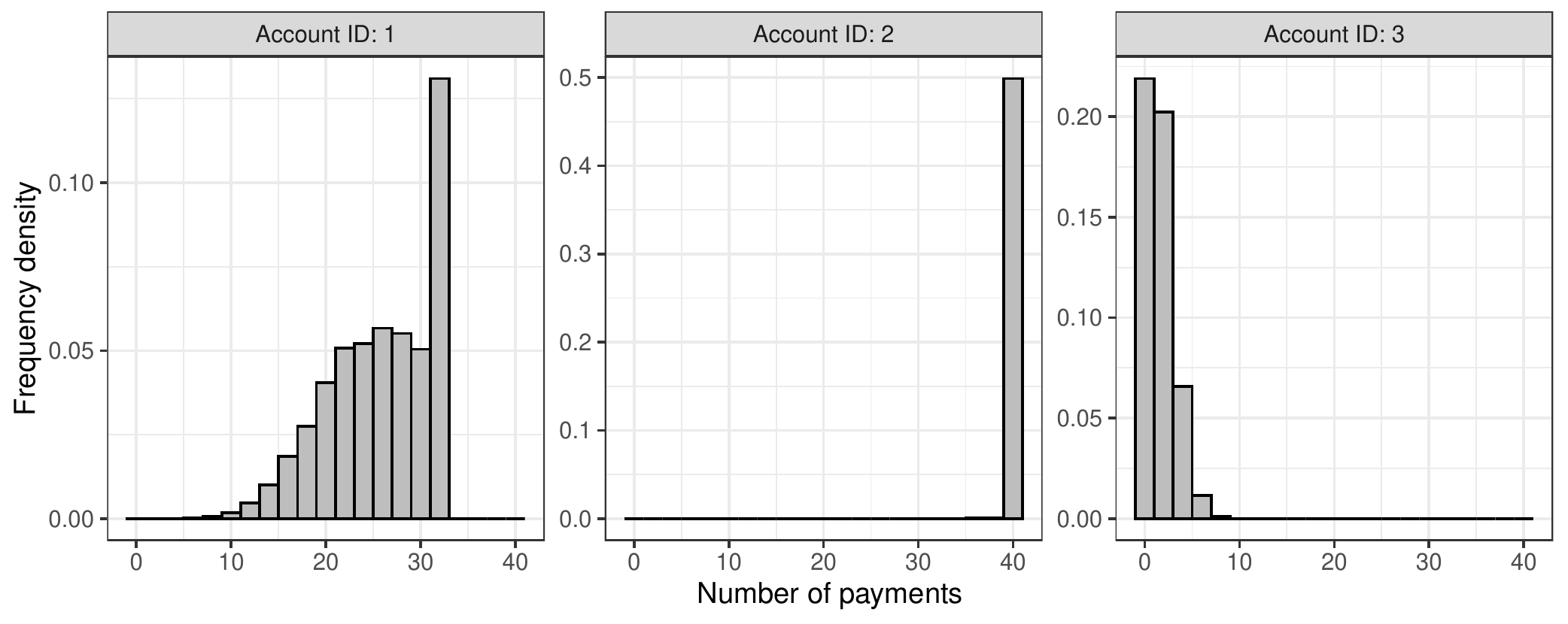}
\caption{Potential balance trajectories (top) and distributions of potential total collections (middle) and number of payments (bottom) for selected accounts obtained via repeated simulation from the representative model. }
\label{fig:heterogeneous-accounts}
\end{figure}

Figure \ref{fig:heterogeneous-accounts} shows the heterogeneity in the behaviour of selected accounts from the simulated population. In particular, Account 139 has a high probability of paying off the complete balance and correspondingly exhibits low uncertainty in the total collections, while the other accounts are more variable. 

\subsubsection{Segment transitions}
\label{sec:transitions}
%Operationally, AGL implements a bespoke customer journey which is influenced by a customer's behaviour and may include escalation in engagement strategies with the customers over time. Each of these strategies come with a set of eligibility criteria, and costs associated with taking the action. In AGL's model, and in the representative model here, accounts can transition between segments, representing a decision to change between predefined strategies.
%The decisions about which accounts to transition is limited by the operational capacity, forcing an account comparison which leads to a source of dependence between the accounts. Accounts are prioritised based on maximising the return on investment.
%The decisions about which accounts to transition are based on a number of factors, including current payment status, ability to pay, and likelihood to recover funds. This decision process leads to those accounts which are eligible for segment transitions to no longer be independent of each other, since the number of transitions is limited by  operational capacity.
%This is an important feature of the model, and as such we describe below a segment transition dynamic within the representative model.

The details of the segment transitions in the representative model are as follows. There are several prespecified transition times, $\tau_m$ ($m=1,\ldots,M$), and in the month corresponding to transition time $\tau_m$ a prespecified number, $n_m$, of accounts is selected to transition to Segment 1. The accounts selected are those with the $n_m$ best credit scores among accounts satisfying the following three conditions: 
\begin{itemize}
    \item[(i)] the account is eligible for transition;
    \item[(ii)] the account is in Segment 3;
    \item[(iii)] the account did not make a payment in the preceding month.
\end{itemize}
Eligibility is decided according to a time-independent indicator covariate \edit{$E_{i}$ which is initialized according to $E_{i} \sim \operatorname{Bernoulli}(0.1)$.} If there are fewer than $n_m$ accounts meeting criteria 1--3, then all accounts meeting the criteria are transitioned but no others. We assume the transition schedule $M=6$,  $(\tau_1,\ldots,\tau_6; n_1,\ldots, n_6)= (6,12,18,24,30,36; 10,10,10,10,10,10)$.  %\slc{Typically is very vague - is this what we do or not? Can we be more specific if not?}

As a result of the segment transitions certain accounts have dependent outcomes. In particular those accounts that are initially in Segment 3 and are eligible to transition have dependent collections, as the decision whether to transition a particular account depends on the payment patterns of the other accounts. We let $\mathcal{D}\subseteq \{1,\ldots, N\}$ denote the indices of these dependent accounts, which we will refer to as the ``dependent block", and $\mathcal{I}= \mathcal{D}^c$ denotes the indices of the independent accounts. \edit{Conditional on segment, all other aspects of performance of accounts are assumed to be independent; this is a feature not only of our model, but also in the construction of AGL's operating model. 

%Dependence between accounts resulting from other sources, such as the macro-economic environment, is a common feature of the literature on probability of default and loss given default modelling (e.g. \cite{huang2011generalized}), but it is not typically incorporated into models for repayment patterns of consumers who have already defaulted (e.g. \cite{so2019debtor}, \cite{thomas2016modelling}). We consider it out of scope for the current paper.}
Dependence between accounts resulting from other sources, such as the macro-economic environment, is a common feature of the literature on probability of default and loss given default modelling (e.g. \cite{huang2011generalized}), but it is not typically incorporated into models for repayment patterns of consumers who have already defaulted (e.g. \cite{so2019debtor}, \cite{thomas2016modelling}). AGL hold the view that while certain economic pressures will impact payment outcomes, defaulted consumers are largely out of step with the wider economy and focus the core of their modelling on individual circumstances. Economic scenarios are overlaid at a later stage to account for economic downturns or upsides. We consider it out of scope for the current paper.}

\subsubsection{Differences with Arrow's commercial model}
%\slc{Paul to check/amend/add to the below}

There are many differences between the representative model and the commercial model, though the similarity is enough to illustrate the application of the methodology. Important differences include the use of real customer data, more appropriate covariates drawn from AGL's extensive feature repository, more detailed simulation of payment amounts with corresponding decisioning models, more granular treatment of different product types, and different criteria for segment transitions.

\subsection{Estimation of expected collections}

To quantify the likely overall outcome, it is of interest to estimate the expected total collections from the population, $\mu = E(\sum_{i=1}^{N} \sum_{t=1}^{84} X_{it})$. This can be done using simulations from the model provided the model is an adequate approximation to the true process, an assumption which is believed to hold approximately for AGL's commercial model. 

\begin{figure}[tpb]
\centering
\includegraphics[width=0.9\linewidth]{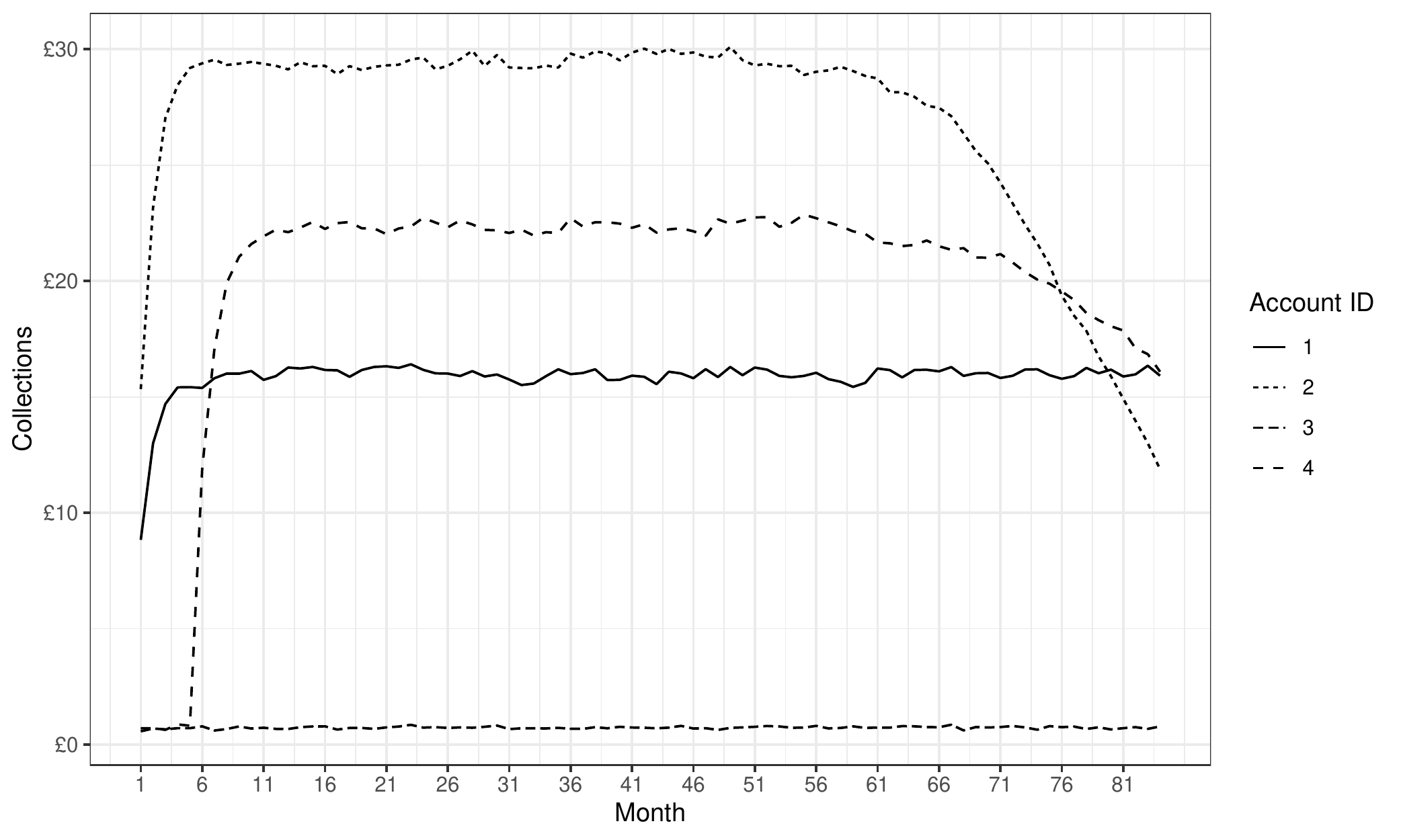}
\caption{Estimated expected collections curves computed for selected representative model accounts using \eqref{eq:estimator-muit}.}
\label{fig:coll-curves}
\end{figure}

The simplest estimation method is to simulate $R$ realisations of the payment behaviour for the entire population from the model. Let $X^{(k)}_{it}$ denote the $k$th realisation of the account level collections in month $t$, and $X^{(k)}_{i}= \sum_{t=1}^{84} X^{(k)}_{it}$ the corresponding realisation of the account level total collections. Then the following Monte Carlo estimator for $\mu$ can be used:
$$
\hat{\mu} = \frac{1}{R}\sum_{k=1}^R \sum_{i=1}^{N} X_i^{(k)}\,. 
$$
This estimator is unbiased for the model-based expectation of the total collections from the population. Similarly the expected total collections in a given time period can be estimated by $\hat{\mu}_{\cdot t} =  \frac{1}{R} \sum_{k=1}^R \sum_{i=1}^{N} X^{(k)}_{it}$, and the expected collections for a specific account in a specific time period can be estimated by 
\begin{equation}
\hat{\mu}_{it} = \frac{1}{R} \sum_{k=1}^R X^{(k)}_{it} \,.
\label{eq:estimator-muit}
\end{equation}
Figure \ref{fig:coll-curves} shows expected collections curves for a selection of accounts in the representative model, estimated using the latter method. 

There are three areas on which we wish to improve on these simple estimators. Firstly, we would like to produce prediction intervals to quantify the uncertainty in the forecast of the population total collections. This is important because the population level total collections may exhibit substantial variability, even though they are relatively predictable compared to the highly volatile individual outcomes owing to the large number of accounts. Accurate estimates of this variability can inform stakeholders about the risk of investment. \edit{We focus on prediction intervals as opposed to other methods of quantifying the uncertainty, up to and including the approximation of the entire distribution of collections, because of its simplicity, both in terms of computational complexity but also for ease of communication to key stakeholders. However other statistical measures of uncertainty could be estimated using methods similar to those presented here, but we note that the computational budget required to give an accurate estimate of the full distribution is prohibitively expensive.}

Secondly, due to the heterogeneity of the accounts, an even spread of computational effort across all of the accounts does not lead to the most accurate estimator possible. We aim to redistribute the computational effort in an optimal way to minimize the variance of the estimator. Lastly, different stakeholders are also often interested in the predicted collections from different identified subsets of the population, referred to as \emph{portfolios}. We may therefore wish to ensure that the quality of the portfolio-level forecasts are preserved, by defining upper bounds on the variances of the corresponding estimators.

\section{Error analysis and optimal number of realisations}\label{sec:Err}

% tw rewrites
We wish to estimate the expected collections as precisely as possible given the available computational budget. 
In practice the population size is very large, typically of the order of millions of accounts, and so  
 restrictions to the computational budget will typically mean that it is only possible to perform a small number of realisations for each individual account within the equal-realisations simulation scheme described in the previous section. This can lead to an undesirably high variance in the estimator for the expected collections. Increasing the global number of realisations per account is not an option due to the large number of accounts and so we require a more sophisticated approach to reduce the estimator variance.

\begin{figure}
\centering
\includegraphics[width=\linewidth]{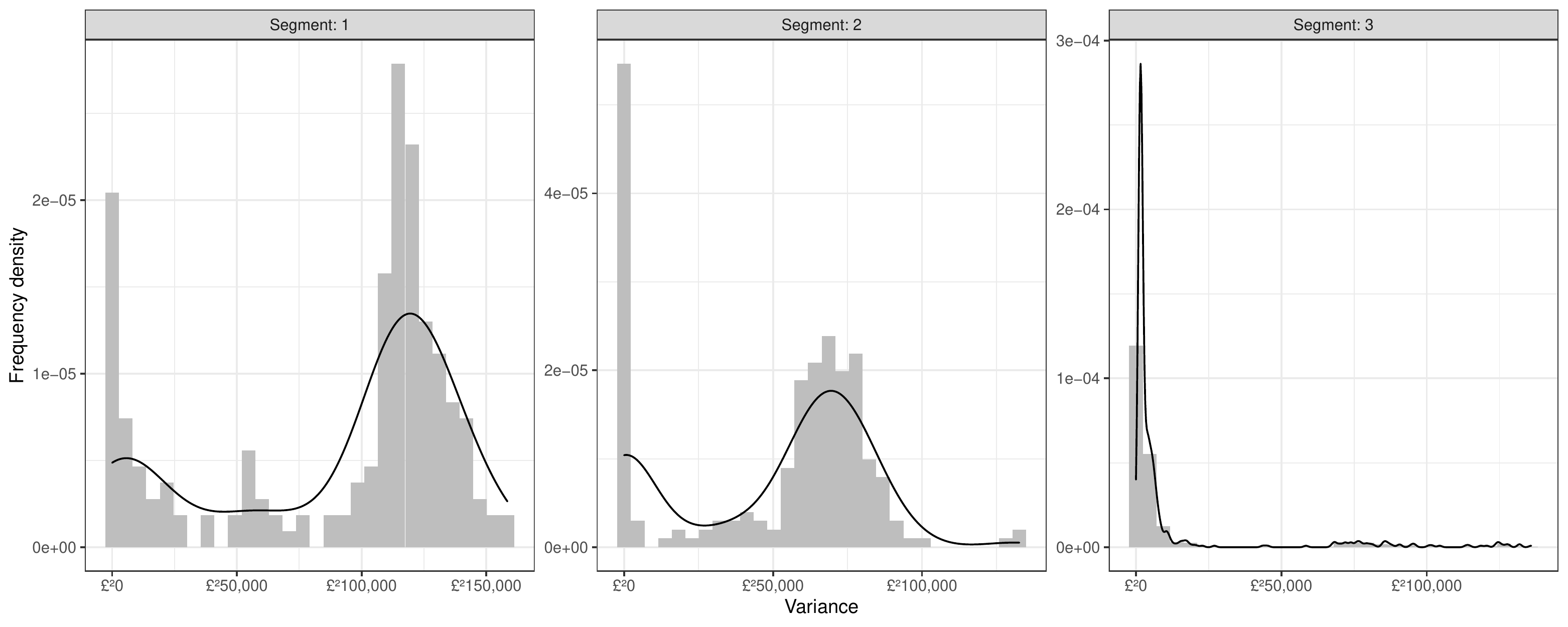}
\caption{Distribution of the account-level variance $\sigma^2_i$ in the simulated population}
\label{fig:var-distribution}
\end{figure}

One way forward can be seen by considering the variance of the estimator of expected total collections in more detail. Note that under the equal-realisations simulation scheme described in the previous section we have
$$
\Var(\hat{\mu}) = \frac{1}{R}\left( \sigma^2_{\mathcal{D}}  + \sum_{i \in \mathcal{I}} \sigma^2_i \right) \,,
$$
where $\sigma^2_i = \Var(X_{i})$ denotes the variance of an individual account and $\sigma^2_{\mathcal{D}}= \Var(\sum_{i \in \mathcal{D}} X_i)$ denotes the variance of the dependent block. Thus an individual account with high variance (i.e. $i$ with large $\sigma^2_i$) will contribute more to the population-level estimator variance than an account with small variance. Moreover, the population does in fact consist of accounts with a wide range of variances, as evidenced for the representative model by Figure \ref{fig:var-distribution}. Thus, with the equal-realisations scheme, in practice the majority of the estimator variance comes from a subset of high-importance accounts. 

The above discussion suggests that it may be better to assign more computational effort to these high importance accounts by performing more simulation realisations for these accounts, and fewer realisations for the low-variance accounts. Let $R_i$ denote the number of realisations for the $i$th account. Note that all accounts in the dependent block must have the same number of realisations due to the fact that these accounts must be simulated together in order to compute the segment transition decisions, hence $R_i \equiv R_{\mathcal{D}}$, $i\in \mathcal{D}$. With this setup we can estimate the expected total collections via
$$
\hat{\mu} 
= \sum_{i=1}^{N} \sum_{k=1}^{R_i} \frac{X^{(k)}_{i} }{R_i}
= \frac{1}{R_{\mathcal{D}}}\sum_{i \in \mathcal{D} } \sum_{k=1}^{R_\mathcal{D}} X^{(k)}_{i} + 
\sum_{i \in \mathcal{I} } \sum_{k=1}^{R_i}  \frac{X^{(k)}_{i} }{R_i} \,,
$$
which is an unbiased estimator for $\mu$ with variance 
\begin{equation}
\Var(\hat{\mu}) = 
\frac{\sigma^2_{\mathcal{D}}}{R_\mathcal{D}} 
+ \sum_{i\in \mathcal{I}} \frac{\sigma^2_i}{R_i} \,.
\label{eq:estimator-variance}
\end{equation}

Assuming for simplicity an equal simulation cost for each realisation of each account, the optimal realisations scheme would be the one which minimizes \eqref{eq:estimator-variance} subject to the budget constraint 
\begin{equation}
C= \sum_{i=1}^{N} R_i=  R_{\mathcal{D}} |\mathcal{D}| + \sum_{i\in \mathcal{I}} R_i\,,
\label{eq:cost-equal}
\end{equation}
where $|\mathcal{D}|$ denotes the number of accounts in the dependent block. 
In order to obtain an analytically tractable solution to this optimization problem we relax the constraint that the $R_i$ should be integers, instead allowing them to be positive real numbers. This parallels the traditional approach in optimal experimental design of using approximate designs rather than exact designs \cite{kiefer1959optimum}. In this case the solution is obtained by finding the unconstrained minimum of the Lagrangian, 
$$
\mathcal{L}(\mathbf{r},  \lambda)
= \frac{ \sigma^2_{\mathcal{D}} }{ R_{\mathcal{D}} }
    +  \sum_{i\in \mathcal{I}}\frac{\sigma^2_i}{R_i} +
    \lambda \left( 
        C - R_{\mathcal{D}} |\mathcal{D}| - \sum_{i\in \mathcal{I}} R_i 
        \right)  \,,
$$
where $\mathbf{r}= (R_{\mathcal{D}}, R_i; i\in\mathcal{I})$. The corresponding optimum is 
\begin{equation}
     R^\ast_{\mathcal{D}} 
 = \frac{ 
     \sigma_{\mathcal{D}} 
    }{ 
        \sqrt{|\mathcal{D}|} 
    }
    \frac{C }{
        (\sqrt{|\mathcal{D}|}\sigma_{\mathcal{D}} +
         \sum_{i\in \mathcal{I}}\sigma_i ) 
    } \,, \qquad
R^\ast_{i} =
    \sigma_i 
 \frac{C }{
        (\sqrt{|\mathcal{D}|}\sigma_{\mathcal{D}} +
         \sum_{i\in \mathcal{I}}\sigma_i ) 
    } \,,\quad 
   (i\in\mathcal{I})\,.
   \label{eq:opt-repeats}
\end{equation}
The assumption of equal simulation costs in \eqref{eq:cost-equal} may not always be a reasonable approximation, for example where there is a large cost associated with the account comparisons involved in the transition decisions. However similar results to \eqref{eq:opt-repeats} could be derived with an alternative cost function. 
%\slc{Added this bit in which could replace the use of $\alpha$ in previous version}
%\tww{thanks! changed wording}

Practical use of the optimal realisations scheme \eqref{eq:opt-repeats} is complicated by the fact that the optimal solution depends on the variances of the independent accounts $\sigma^2_i$ ($i\in \mathcal{I})$ and the variance of the dependent block, $\sigma^2_{\mathcal{D}}$. 
These quantities are unknown before the simulations have been carried out, thus to obtain a feasible realisations scheme we propose pre-estimating the variances as follows. 
Firstly, for the independent accounts we suggest using a pre-trained Gaussian process emulator to predict $\sigma^2_i$ from the account covariates $\mathbf{Z}_{i1}$ and the preceding month payment indicator $Y_{i0}$; see Section \ref{sec:GP-predict-variance} for full details of the construction of such an emulator. 
Secondly, for $\sigma^2_{\mathcal{D}}$ we suppose that we have available a number of pilot realisations of the dependent accounts. 
The sample variance of the total collections from these realisations is used to give a pilot estimate, $\hat{\sigma}^2_{\mathcal{D}}$. 

Given the variance pre-estimates, an efficient realisations scheme is obtained by plugging the variance pre-estimates in to \eqref{eq:opt-repeats}. 
The realisation numbers resulting from the formulae are then rounded to the nearest integer, aside from values in $(0,1)$ which are rounded up to unity. 
More efficient rounding schemes (e.g. similar to those used for approximate designs \cite{pukelsheim1992efficient}) could be explored, but are outside the scope of this paper.

Above we used a different pre-estimation method for the dependent and independent accounts. 
The reason for this is that $\sigma^2_{\mathcal{D}}$ is a function of the the vector $(\mathbf{Z}_{i1}, Y_{i0}; i\in \mathcal{D})$ of covariates and initial payment indicators of all accounts in the dependent block, and it is unlikely to be feasible to construct an accurate and computationally efficient Gaussian process emulator on such a high dimensional space. 
Moreover the size of the dependent block will likely vary across different forecasting rounds, so the emulator would need to be retrained on each occasion at high cost. 
This problem is sidestepped by using the straightforward pilot simulation approach above. \edit{The reason for changes in the dependent block is due to segment transitions of a subset of the accounts between forecasting rounds.}

Another notable aspect of the method above is the treatment of the pilot realisations. First, the additional cost associated with the pilot realisations is ignored in the cost function \eqref{eq:cost-equal}. This is reasonable because AGL run repeated forecasts for each forecasting round, and so the overhead costs can easily be absorbed. 
Second, the pilot realisations are used only to estimate the number of realisations; they are not used directly in constructing the estimator $\hat{\mu}$.
The alternative, which may be more desirable if the computational budget is extremely limited, would be to include the additional cost of the pilot realisations in \eqref{eq:cost-equal} and to include the pilot realisations in $\hat{\mu}$. 
However this would introduce dependence between $X_{\mathcal{D}}$ and $X_{i}$ ($i\in\mathcal{I}$) and invalidate the variance formula \eqref{eq:estimator-variance}, necessitating a more complicated analysis.

\begin{figure}[tbp]
\centering
\includegraphics[width=0.7\linewidth]{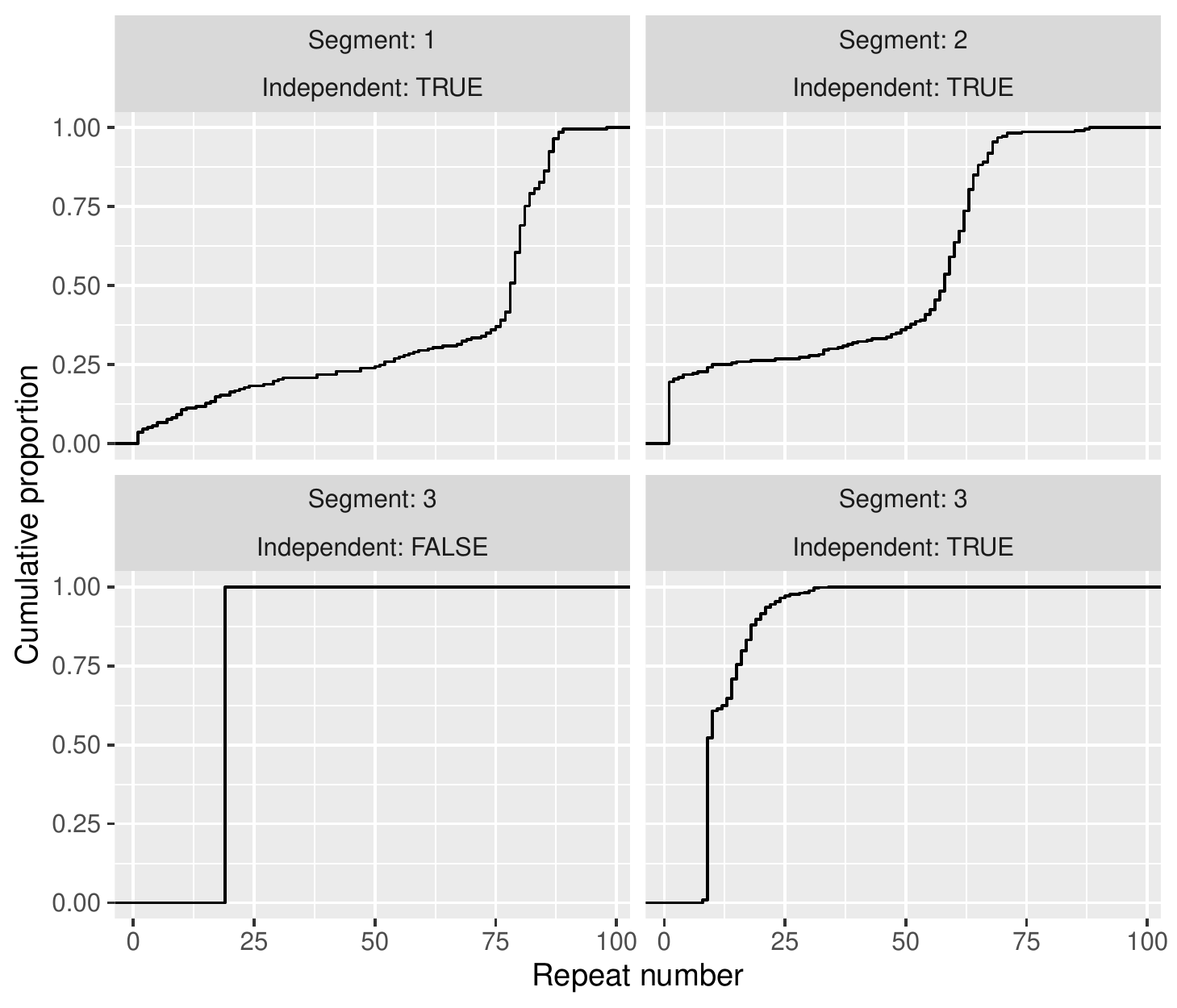}
\caption{Distribution of the realisation number $R_i$ across the population when using \eqref{eq:opt-repeats} with variance pre-estimation. The vertical axis gives the empirical cumulative distribution function for $R_i$, i.e.~the proportion of accounts with a realisation number less than or equal to the value shown on the horizontal axis.}
\label{fig:repeat-number-distribution}
\end{figure}

Figure \ref{fig:repeat-number-distribution} shows the distribution of realisation numbers that results from applying the above optimization procedure to the representative model with $N=1000$ accounts. We see that there is a wide range of realisation numbers, and that the computational effort varies substantially across segments. In particular, for the accounts in Segment 3 the dependent accounts receive 19 realisations each and the independent accounts receive a median of 9 realisations. In contrast, in Segments 1 and 2 the median realisation numbers are 78 and 58 respectively, with some realisation numbers as high as 99 in Segment 1. Around 4\% of accounts in Segment 1 and 20\% in Segement 2 receive just one realisation. Extensive simulation shows that the effect of this optimization is to reduce $\Var(\hat{\mu})$ from $1.22\times 10^6$ to $8.17 \times 10^5$, a reduction of 33\%. 

%\ptr{Is the cdf within segment figure the best representation for this? Segment 3 independent block has by far the lowest average repeat numbers, and it is the distribution of repeat numbers across all segments which give us the savings, rather than within a segment. The MC error reduction is best when a larger proportion of accounts are in Segment 3 Ind.}

%\tww{the 'extensive simulation' doesn't resample the pilot runs, just the follow-up, so it probably underestimates the variance of optimized approach...}
%\slc{We should redo this, so that the pilot runs are overhead costs, as in the (new) text.}

\begin{figure}
\centering
\includegraphics[width=0.9\linewidth]{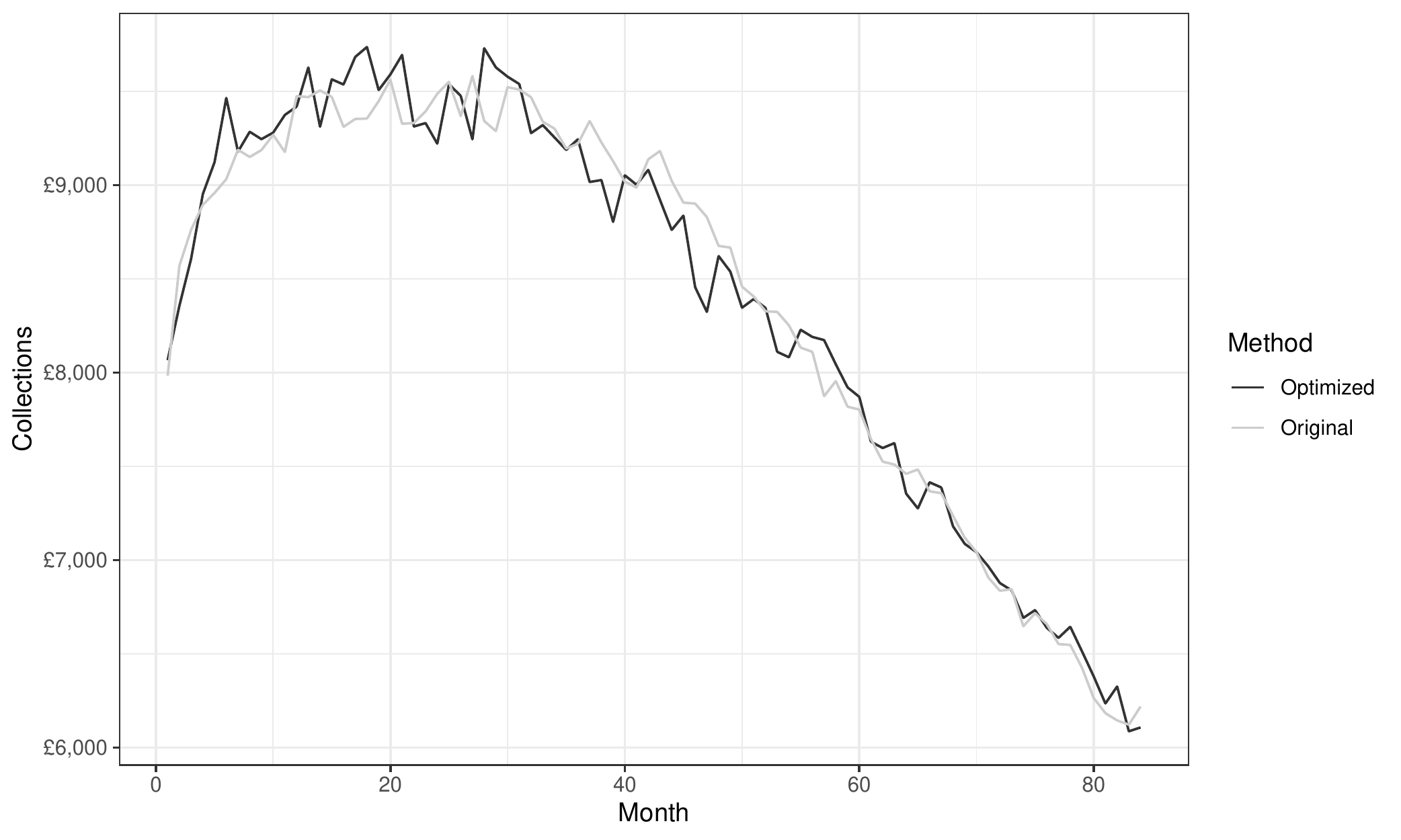}
\caption{Estimated population-level collections curves, using both the equal realisations simulation setup and the optimized setup \eqref{eq:opt-repeats} with variance pre-estimation.}
\label{fig:population-collections-curves}
\end{figure}

Figure \ref{fig:population-collections-curves} shows the estimates of the expected collections curves using both the equal realisations and optimized realisations methodologies. Interestingly, the curve is noisier with the optimized realisation methodology. This is somewhat unexpected given that this method has lower variance for total collections up to 84 months. However, a potential explanation is that while the variance of the total collections is lower for the optimized method, the variances at individual time periods may be higher for some times. For example if a payment is certain to pay their full balance and the only uncertainty is in the timing then the realisation scheme from \eqref{eq:opt-repeats} will allocate fewer realisations to this account, increasing the variance in individual time periods. 

%\slc{Nice explanation!}

\section{Uncertainty quantification}\label{sec:UQ}

The point estimate of the expected collections does not give a complete picture, since the collections vary between realisations. It is also important for stakeholders to be able to quantify the possible variation in the collections, enabling investors to balance risk and
reward. We are interested to know how much the true collections are likely to
vary from the forecasted mean; this can be quantified using a prediction interval.

\subsection{Prediction intervals for total collections}

One possible way to define a prediction interval for the true total collections $X$  with approximate coverage probability $p$ (e.g. $p=0.95$) is to use the end-points
\begin{equation}
    \hat{\mu} \pm z_{(1+p)/2} \sqrt{\hat{\sigma}^2_{(X-\hat{\mu})}} \,, 
    \label{eq:prediction-interval}
\end{equation}
where $\hat{\sigma}^2_{(X-\hat{\mu})}$ denotes an appropriate estimate of the variance, $\sigma^2_{(X-\hat{\mu})}$, of the prediction error $X-\hat{\mu}$ and $z_q$ denotes the $q$-quantile of a standard normal distribution. There are two main questions about such an interval: (i) how to construct the variance estimate $\hat{\sigma}^2_{(X-\hat{\mu})}$, and (ii) whether the choice of a standard normal quantile is appropriate. We suggest two possible methods for (i), and for (ii) we identify two sets of asymptotic conditions under which the interval's coverage correctly converges to the nominal value. 

To form the estimate $\hat{\sigma}^2_{(X-\hat{\mu})}$ in Equation \eqref{eq:prediction-interval}, we will substitute estimates $\hat{\sigma}^2_{\mathcal{D}}$ and $\hat{\sigma}^2_i$ of the variances $\sigma^2_{\mathcal{D}}$ and $\sigma^2_i$ into the expression\edit{
\begin{equation}
\Var(X-\hat{\mu})
= \sigma^2_{\mathcal{D}}(1+1/R_\mathcal{D}) + 
  \sum_{i\in\mathcal{I}} \sigma^2_{i} (1+1/R_i) \,.
\label{eq:prediction-error-variance}    
\end{equation}}
Here $\sigma^2$ denotes the natural variance of the collections of the accounts/dependent block, with the additional \edit{$\sigma^2/R$} terms coming from the variance of the Monte Carlo estimates of the mean collections.
Equation \eqref{eq:prediction-error-variance} holds under the assumption that the true collections are an independent realisation from the same model, an assumption which should be justified at least approximately for the commercial model. Two sources of
uncertainty are incorporated into \eqref{eq:prediction-error-variance}: the natural variability of the collections due to the
stochastic nature of the model and the variance of the estimator of
the expected collections. 

To estimate $\sigma^2_{\mathcal{D}}$ there is little choice but to use the sample variance of the realisations of the total collections from the dependent accounts, i.e.
$
\hat{\sigma}^2_{\mathcal{D}} =
\frac{1}{R_{\mathcal{D}}-1} 
\sum_{k=1 }^{R_{\mathcal{D}}}
(X_{\mathcal{D}}^{(k)} - \bar{X}_{\mathcal{D}})^2
$, 
where 
$X^{(k)}_{\mathcal{D}} = \sum_{i\in \mathcal{D} } X^{(k)}_i$ 
denotes the $k$th simulated realisation of the total collections from the dependent block, and $\bar{X}_{\mathcal{D}} = \frac{1}{R_{\mathcal{D}}} \sum_{k=1}^{R_{\mathcal{D}}} X^{(k)}_{\mathcal{D}}$ the sample mean of these realisations. Meanwhile, for $\hat{\sigma}^2_i$ there are two choices of method:-
\begin{itemize} 
\item[(M1)] the sample variance of the realisations of the $i$th account, i.e. $\hat{\sigma}^2_i = \frac{1}{R_i-1}\sum_{k=1}^{R_i} (X_i^{(k)} - \bar{X}_i)^2$; or 
\item[(M2)] the output of the Gaussian process emulator for the variance of an individual independent account, as discussed in Section \ref{sec:GP-predict-variance}.
\end{itemize}
We will show that Method 1 above gives an asymptotically valid interval, but it requires $R_i \geq 2$ for all accounts. Method 2 should also give an interval with approximately correct coverage provided the GP variance model is sufficiently accurate, but there are no theoretical guarantees as to the degree of accuracy achieved. 

\begin{figure}[tbp]
\centering
\includegraphics[width=0.65\linewidth]{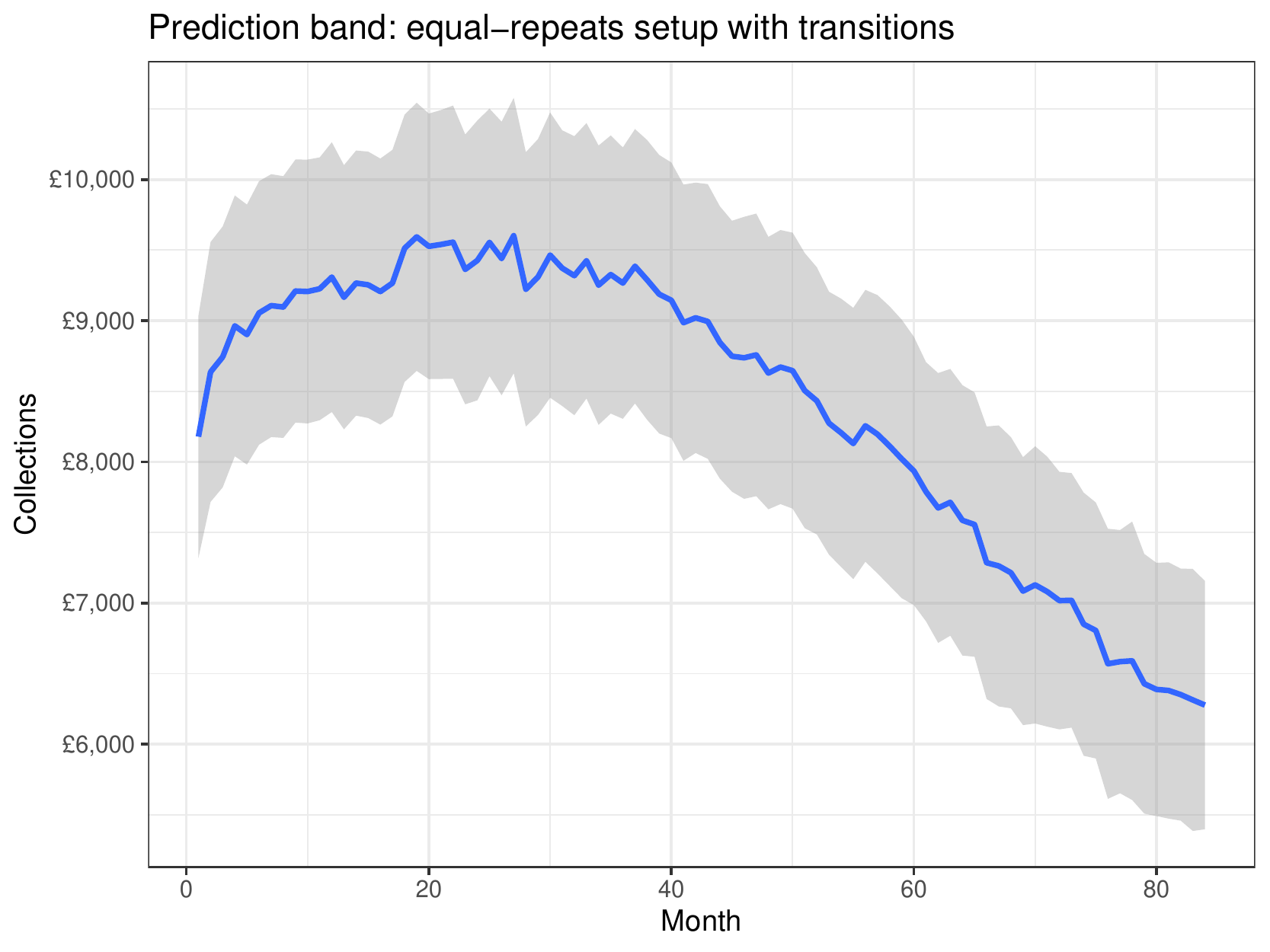}
\caption{95\% prediction bands for collections in each month, using equal realisations setup.}
\label{fig:prediction-bands}
\end{figure}

A similar approach can be used to define prediction intervals for the collections in a given time period, by replacing $\hat{\mu}$, $\hat{\sigma}^2_{\mathcal{D}}$, $\hat{\sigma}^2_i$ with estimates of the relevant time-specific quantities. Figure \ref{fig:prediction-bands} shows the use of this approach to quantify uncertainty in the forecasts using the equal realisations simulation setup. It is clear that there is substantial uncertainty which is roughly constant over time, but the width of the interval is of a smaller order of magnitude than the expected collections. For the optimized realisations setup the prediction band could not be computed as some accounts have $R_i<2$, precluding the use of the sample variance to estimate $\sigma^2_{it} = \Var(X_{it})$, needed in the standard error. \edit{Prediction of the time-specific $\sigma^2_{it}$ with Gaussian processes would require substantial changes to the methodology in Section \ref{sec:GP-predict-variance}, which are outside the scope of the present paper.}

\subsection{Asymptotic results}
In order to study theoretical properties of the interval \eqref{eq:prediction-interval}, we consider an asymptotic scenario in which the number of independent accounts $|\mathcal{I}| \to \infty$. Our results are contingent on a number of regularity conditions which control how the population is assumed to grow and which require some further notation. Let $I\sim \text{Unif}(\mathcal{I})$ denote the index of an account selected randomly from the population of independent accounts, and let $\bar{\sigma}^2 = \E(\sigma^2_I) = \frac{1}{|\mathcal{I}|} \sum_{i\in \mathcal{I}} \sigma^2_i$ and $\gamma^2= \Var(\sigma^2_I)/\bar{\sigma}^4$ denote respectively the expectation and squared coefficient of variation of the variance of this account. Then the regularity conditions are:
\begin{enumerate}
\item[(R1)] $R_i \geq 2$, $i\in\mathcal{I}$, with the $R_i$ fixed. 
\item[(R2)] $\kappa_i \leq \kappa_\text{max}$, $i\in\mathcal{I}$, where $\kappa_i$ denotes the kurtosis of $X_i$ and the bound $\kappa_\text{max}$ is independent of $|\mathcal{I}|$. 
\item[(R3)] $\gamma^2$ is bounded as $|\mathcal{I}|\to \infty$.
\item[(R4)] $\kappa_{\mathcal{D}} = O(|\mathcal{I}|\bar{\sigma}^2)$, where $\kappa_{\mathcal{D}}$ denotes the kurtosis of $X_{\mathcal{D}}$.
   \item[(R5a)] $\sigma^2_{\mathcal{D}} = o(|\mathcal{I}|\bar{\sigma}^2)$.
\end{enumerate}
\edit{Above, for functions $f(n),g(n)\geq 0$, $n\in\mathbb{N}$, the notation $f(n)= O(g(n))$ means that there exists $n'$ and $K$ such that $|f(n)|/g(n)\leq K$ for all $n>n'$. Moreover, $f(n)=o(g(n))$ means that $f(n)/g(n)\to 0$ as $n\to \infty$.}

Condition (R5a) corresponds to an assumption that the variance of the dependent block is asymptotically of smaller magnitude than that of the independent block. This can be avoided by assuming approximate normality of the contribution from the dependent accounts, and a large number of simulated realisations of the dependent block. Empirical results suggest that the normal approximation for the dependent block is reasonable, but it is difficult to establish theoretically. This alternative assumption is formalized as follows: 
\begin{enumerate}
 \item[(R5b)] 
 \begin{itemize}
     \item[(i)] $\sigma^2_{\mathcal{D}} \asymp \beta_1 |\mathcal{I}| \bar{\sigma}^2$ for fixed $\beta_1>0$;
     \item[(ii)]   \edit{$\Var(X_{\mathcal{I}}- \hat{\mu}_{\mathcal{I}})\asymp \beta_2  |\mathcal{I}| \bar{\sigma}^2$}, for fixed $\beta_2 \in [1,2]$;
     \item[(iii)]  $R_\mathcal{D}\to \infty$; 
     \item[(iv)] $X_{\mathcal{D}}$ is asymptotically approximately normal, i.e. \edit{$$
     \frac{(X_{\mathcal{D}} - \mu_{\mathcal{D}})}{\sqrt{\Var(X_{\mathcal{D}}- \mu_{\mathcal{D}}})}\to N(0,1)$$} in distribution as the population size tends to infinity, \edit{where $\mu_{\mathcal{D}}=\E(X_{\mathcal{D}})$}.
 \end{itemize} 
\end{enumerate}
Above we write $f(n) \asymp g(n)$ if $f(n)/g(n) \to 1$ as $n\to \infty$. Note that (R5b)(i) states that the variance of the dependent block is of comparable magnitude to that of the independent block, while (R5b)(ii) is a rather weak additional constraint given the previous conditions already imply that 
\edit{$ |\mathcal{I}| \bar{\sigma}^2 \leq 
Var(X_{\mathcal{I}} - \hat\mu_{\mathcal{I}})
%\sigma^2_{(X_{\mathcal{I}} - \hat\mu_{\mathcal{I}})} 
\leq 2 |\mathcal{I}| \bar{\sigma}^2$.}

\begin{theorem}\label{thm1}
Suppose that $|\mathcal{I}|\to \infty$ with regularity conditions (R1)--(R4) and that either (R5a) or (R5b) holds. Then the interval \eqref{eq:prediction-interval} with variance estimated according to Method 1 is an asymptotically valid $100p\%$ prediction interval for the total collections.
\end{theorem}

\edit{The proof and supporting lemmas are presented in Appendix \ref{sec:Proof1}.}

\subsection{Coverage study}

To assess and compare the performance of the prediction intervals from this section we carried out a simulation study using the representative model (see Table \ref{tab:coverage-results}). For the GP-based intervals the realisation numbers were optimized using the method of Section \ref{sec:Err}. The results show that the coverage, i.e. the probability that the prediction interval contains the true realised collections, is close to the nominal level of 95\% even for quite small population sizes such as $N=100$, and so the asymptotic results appear highly robust in practical situations. In the simulation studies we have undertaken the \edit{mean simulated} length of the optimized prediction intervals is   shorter than those using equal realisation numbers, however the magnitude of the reduction is quite small due to the irreducible underlying variability of the accounts. The absolute uncertainty in the predictions increases with $N$, but this is to be anticipated as the expected total collections also grow with the population size. If we instead work with the relative uncertainty, meaning the width of the prediction interval divided by its central value, then we see that the uncertainty decreases as $N$ increases.
%\slc{Could display the same statistic normalising for population size to show the decrease?}  

\begin{table}
    \caption{\label{tab:coverage-results}Prediction interval performance. Performance metrics are estimated using 1,000 independent repetitions of the entire procedure. `Relative uncertainty' denotes the mean of the interval width divided by the interval mid-point.}
    \centering
    \begin{tabular}{p{1.8cm}llp{1.8cm}p{1.8cm}}Population size, $N$    &   Interval method& Coverage  & \edit{Mean length} & Relative uncertainty \\
    \hline 
    100 & GP-free, equal &  95.9\%  &  7772.6 & 12.4\% \\
    100 & GP-based, optimized & 96.0\% & 7696.6 & 12.3\% \\
    250 & GP-free, equal & 95.5\%    & 11865.0& 6.22\% \\
    250 & GP-based, optimized & 95.2\% & 11768.8 & 6.17\%\\
    1000 & GP-free, equal & 94.2\% & 22644.8 & 3.42\%\\ 
    1000 & GP-based, optimized & 95.0\% & 22608.0 & 3.42\% 
    \end{tabular}
\end{table}

%N=10
%length_original  length_optimized  covered_original covered_optimized 
%         2434.621          2405.063             0.941             0.944 
%
%N=50
%length_original  length_optimized  covered_original covered_optimized 
%         4528.618          4507.059             0.953             0.943 
%
%N=1000
%  length_original  length_optimized  covered_original covered_optimized 
%        21219.186         21060.183             0.959             0.958 

The fact that the central limit asymptotics are highly accurate even for small population sizes is surprising. One potential explanation is that perhaps, owing to the large number of time points involved,  the total collections at the level of a typical individual account level may themselves be close to normally distributed (cf.~the Markov chain central limit theorem \cite{tierney1994markov}). Aggregating approximately normal individual collections will lead to approximately normal total population collections even for small populations. Of course, approximate normality will not hold at the individual level for high kurtosis accounts, but smaller simulated populations are less likely to contain such accounts. 

Note that to estimate the coverage probability we also required realisations of the `true' collections; in our study these were simulated independently from the representative model using $R_i=1$ for all individuals. This corresponds to an assumption that the model used to find prediction intervals accurately reflects the true process we are trying to predict, an assumption that obviously does not hold for the representative model. Nonetheless the results indicate it is reasonable to suppose that the prediction intervals obtained by combining the above methodology with AGL's more accurate commercial model will be well calibrated.

\section{Protection of portfolio-level forecasts}
\label{sec:protect}
Often the population is partitioned into disjoint subsets called \emph{portfolios} with different stakeholders. Thus, while we wish to minimize the variance of the overall population-level forecast, it is often also of interest to protect the variance of the portfolio-level forecasts. In this section we develop methodology to address this constrained optimization problem. 

First we establish notation. Let $\mathcal{P}_j \subseteq \{1,\ldots,N\}$ denote the indices of the accounts in the $j$th portfolio ($j=1,\ldots,J$) within the population. Further let $\mathcal{D}_j \subseteq \mathcal{P}_j$ and $\mathcal{I}_j = \mathcal{P}_j \backslash \mathcal{D}_j$ denote respectively the indices of the dependent accounts and independent accounts within the $j$th portfolio, and $\mathcal{I}= \cup_j \mathcal{I}_j$ the complete set of independent accounts. Note that we assume there is no dependence across different portfolios, which is indeed the case if the segment transitions are applied on a per-portfolio basis. Let $R_i$ denote the number of realisations of the $i$th account in the population. Within each dependent block all accounts must be simulated together and so they have equal realisation numbers, thus $R_i \equiv r_j$ for all $i \in \mathcal{D}_j$. 

The expected collections in the $j$th portfolio can be estimated unbiasedly via
$
\hat{\mu}_j 
= \sum_{i \in \mathcal{P}_j} \sum_{k=1}^{R_i} X^{(k)}_i/ R_i 
= \sum_{i \in \mathcal{D}_j} \sum_{k=1}^{r_j} X^{(k)}_i/r_j 
+ \sum_{i \in \mathcal{I}_j} \sum_{k=1}^{R_i} X^{(k)}_i / R_i 
$,
with variance
$$
\Var{\hat{\mu}_j} = \frac{\sigma^2_{\mathcal{D},j}}{r_j} + \sum_{i \in \mathcal{I}_j} \frac{\sigma^2_i}{R_i}\,,  
$$
where $\sigma^2_{\mathcal{D},j}= \Var(\sum_{i\in \mathcal{D}_j} X_i )$ denotes the variance of the total collections for the accounts in the $j$th dependent block. In addition the total expected collections across the population can be estimated via $\hat{\mu} = \sum_j \hat{\mu}_j$ with variance $\Var(\hat{\mu}_j) = \sum_j \Var(\hat{\mu}_j)$.

Our constrained optimization problem can thus be formulated as follows. We wish to find the optimal realisation numbers  $\mathbf{r}^\ast=(r^\ast_1,\ldots,r^\ast_J, R^\ast_i; i\in \mathcal{I} )$ that minimize $\Var(\hat{\mu})$ subject to the portfolio-level variance constraints $\Var(\hat{\mu}_j) \leq V_j$ and the cost constraint $\sum_j r_j |\mathcal{D}_j| + \sum_{i \in \mathcal{I}} R_i = C$. Similar to previous sections we relax this problem by allowing the $r_j, R_i$ to be positive real numbers, not just integers. By the results in the preceding paragraph the corresponding Lagrangian is
\begin{eqnarray*}
    \mathcal{L}(\mathbf{r}, \lambda, \boldsymbol{\delta} )& = &
    \sum_{j=1}^J \left( 
        \frac{\sigma^2_{\mathcal{D},j}}{r_j} 
        + \sum_{i \in \mathcal{I}_j} \frac{\sigma^2_i}{R_i} 
    \right) + 
    \lambda \left( 
        \sum_{j=1}^J r_j |\mathcal{D}_j| 
        + \sum_{j=1}^J \sum_{ i \in \mathcal{I}_j} R_i  
        - C 
        \right) \\& & +
    \sum_{j=1}^{J} \delta_j 
        \left( 
             \frac{\sigma^2_{\mathcal{D},j}}{r_j} 
             + \sum_{i \in \mathcal{I}_j} \frac{\sigma^2_i }{R_i} 
             -  V_j
        \right) \,,
\end{eqnarray*}
and the solution to the optimization problem must satisfy the Karush-Kuhn-Tucker conditions:
\begin{itemize}
    \item stationarity, i.e. 
    $\frac{\partial \mathcal{L}}{\partial r_j}=0$ ($j=1,\ldots,J$) and
    $\frac{\partial\mathcal{L}}{\partial R_i}=0$ ($i\in \mathcal{I}$);
    \item primal feasibility, i.e. the cost and variance constraints are satisfied;
    \item dual feasibility, i.e. $\delta_j \geq 0$;
    \item complementary slackness, i.e. for all $j$ either $\delta_j=0$ or the corresponding variance inequality constraint holds with equality. In the latter case we say the constraint is `active'.
\end{itemize}

Let $\mathcal{B} = \{ j : \Var(\hat{\mu}_j) = V_j \}$ denote the indices of the active inequality constraints, noting that $\delta_j=0$ for $j\not\in \mathcal{B}$. Given $\mathcal{B}$ the solution of the stationarity equations is given by the following:
\begin{align}
\intertext{\underline{Accounts in an actively constrained portfolio ($i\in\mathcal{P}_j$, $j\in \mathcal{B}$):}} 
    R^\ast_i &= 
    \sigma_i 
    \left(
    \frac{ 
        \sigma_{\mathcal{D},j} \sqrt{|\mathcal{D}_j|} 
        + \sum_{i'\in\mathcal{I}_j} \sigma_{i'} 
        }{V_j} 
    \right)\,, \quad i \in \mathcal{I}_j , j \in \mathcal{B}\,,  \label{eq:Repeats-constrained-indep} \\[1ex]
    R^\ast_i &= r^\ast_j = 
    \frac{ \sigma_{\mathcal{D},j} }{ \sqrt{|\mathcal{D}_j|} } 
    \left(
      \frac{
        \sigma_{\mathcal{D},j} \sqrt{|\mathcal{D}_j|} 
        + \sum_{i' \in \mathcal{I}_j} \sigma_{i'} 
      }{V_j }
    \right) \,, \quad  i \in \mathcal{D}_j, j \in \mathcal{B}\,, 
    \label{eq:Repeats-constrained-dep} 
\intertext{\underline{Accounts in a non-actively constrained portfolio ($i \in \mathcal{P}_j$, $j \in \mathcal{B}^c$)}:}
    R^\ast_i &= 
     \sigma_i \frac{ C^{\mathcal{B}}_{\text{Rem}} }{ d^\mathcal{B} } 
     \,, 
     \quad  i \in \mathcal{I}_j, j \in \mathcal{B}^c \,,
     \label{eq:Repeats-unconstrained-indep}
     \\[1ex]
    R^\ast_i &= r^\ast_j = 
    \frac{
        \sigma_{\mathcal{D},j} 
    }{
        \sqrt{|\mathcal{D}_j|}   
    }  
    \frac{ C^{\mathcal{B}}_{\text{Rem}}  }{ d^{\mathcal{B}} } \,,
    \qquad  i \in \mathcal{D}_j, j \in \mathcal{B}^c \,,
    \label{eq:Repeats-unconstrained-dep}
\end{align}
where above
$$
C^\mathcal{B}_\text{Rem} 
= 
C 
- \sum_{j\in \mathcal{B}} \sum_{i' \in \mathcal{I}_j} R^\ast_{i'}
- \sum_{j \in \mathcal{B} } |\mathcal{D}_j| r^\ast_j 
$$ 
denotes the computational budget remaining after allocating the appropriate amount of budget to the actively constrained portfolios according to \eqref{eq:Repeats-constrained-indep} and \eqref{eq:Repeats-constrained-dep}, and the denominator is a normalising constant given by 
$$
d^\mathcal{B}= 
\sum_{j\in \mathcal{B}^c} \left(
    \sigma_{\mathcal{D},j} \sqrt{|\mathcal{D}_j|} 
    + \sum_{i' \in \mathcal{I}_j} \sigma_{i'} 
\right)\,.
$$
The set $\mathcal{B}$ can be determined numerically using Algorithm \ref{alg:active-set}\edit{. This algorithm  can be shown to converge to the global optimum active set given mild regularity conditions provided it is supplied with the true variances as inputs (see Appendix \ref{sec:algo-convergence}). In practice, however, the required variances must be pre-estimated using a method similar to Section \ref{sec:Err}, and numerical errors in the estimates may mean that the constraints are met only approximately by the estimated solution.}

\begin{Algorithm}[hbtp]
\begin{algorithmic}[1]
\STATE{initialize the set of active constraints as $\mathcal{B}= \emptyset$}
\REPEAT 
\STATE{using the current active set, compute the optimal realisation allocation $(r^\ast_1,\ldots, r^\ast_J, R^\ast_i; i\in \mathcal{I})$ from Equations}  \eqref{eq:Repeats-constrained-indep}--\eqref{eq:Repeats-unconstrained-dep} 
\FOR{$j=1:J$}
\IF{ the $j$th variance constraint is not satisfied, i.e. $\Var(\hat{\mu}_j) > V_j$ }
  \STATE{include $j$ in the active set}, i.e. $\mathcal{B} \gets \mathcal{B} \cup \{j\}$
\ENDIF
\ENDFOR
\UNTIL{a complete pass of the for loop results in no additions to the active set}
\RETURN{the active set $\mathcal{B}$ and the corresponding optimal allocation}
\end{algorithmic}
\caption{Solution of the inequality-constrained optimization problem \label{alg:active-set}}
\end{Algorithm}

To illustrate the effect of this methodology we consider a setup with a population of $N=1000$ accounts divided into two portfolios. The portfolios were defined by sampling an indicator  with a 99\% probability of belonging to Portfolio 1 and a 1\% probability of belonging to Portfolio 2, independent of other accounts and the other variables. If no portfolio-level variance protection is applied then optimization of the realisation numbers yields a variances of $868.1^2$ and $119.3^2$ for the total collections in Portfolios 1 and 2, with Portfolio 2 having ten times greater variation relative to the expected collections. Applying portfolio protection with $(V_1,V_2)= (1000^2, 50^2)$ leads to a realisation distribution where much more computational effort is applied to accounts in Portfolio 2, in particular with an average realisation number of 84.8 compared to 29.2 for Portfolio 1. The effect of this redistribution of effort is to 
change the portfolio-level variances from $(868.1^2,119.3^2)$ to  $(899.6^2, 48.75^2)$, which meets the variance constraint. On other runs of the same code \edit{the variance in Portfolio 2 did not meet the constraint, due to differences between the pre-estimated values of $\sigma^2_{\mathcal{D},j}$ and $\sigma^2_i$ and their true values. However the variance was nonetheless always close to the desired value and much lower than with the previous approach}.  

%\tww{What about pre-estimation of the variances/ lower bounds on the number of repeats of the dependent accounts?}
%\tww{Again the simulations do not take into account the variance due to the pre estimation step}
%\slc{As above, we should rerun these simulations with the pre-estimation being an overhead rather than these simulations being used in the estimators.}
%\slc{My understanding is that the simulations did NOT include the pre-estimation runs, so I think this is OK now?}

\section{Predicting the variance of independent accounts using Gaussian process emulators}
\label{sec:GP-predict-variance}

The variance of an independent account can be considered as a function of the account covariates, i.e. $\sigma^2_i=\Var(X_i) = v(B_{i1}, C_{i}, S_{i1}, Y_{i0})$, $i\in \mathcal{I}$. In this section we train a Gaussian process emulator \cite{sacks1989design,bastos2009diagnostics}  to approximate the function $v(B, C, S, Y_0)$. Once the emulator has been trained it can be used to rapidly predict the variance of new accounts without further simulation from the model, facilitating optimization of the computational budget using Equation \eqref{eq:opt-repeats} and uncertainty quantification as in Section \ref{sec:UQ}. Note that the function $v$ does not depend on the eligibility indicator $E_{i}$, as the collections are conditionally independent of eligibility given segment.

\begin{figure}
    \centering
    \includegraphics[width=0.9\linewidth]{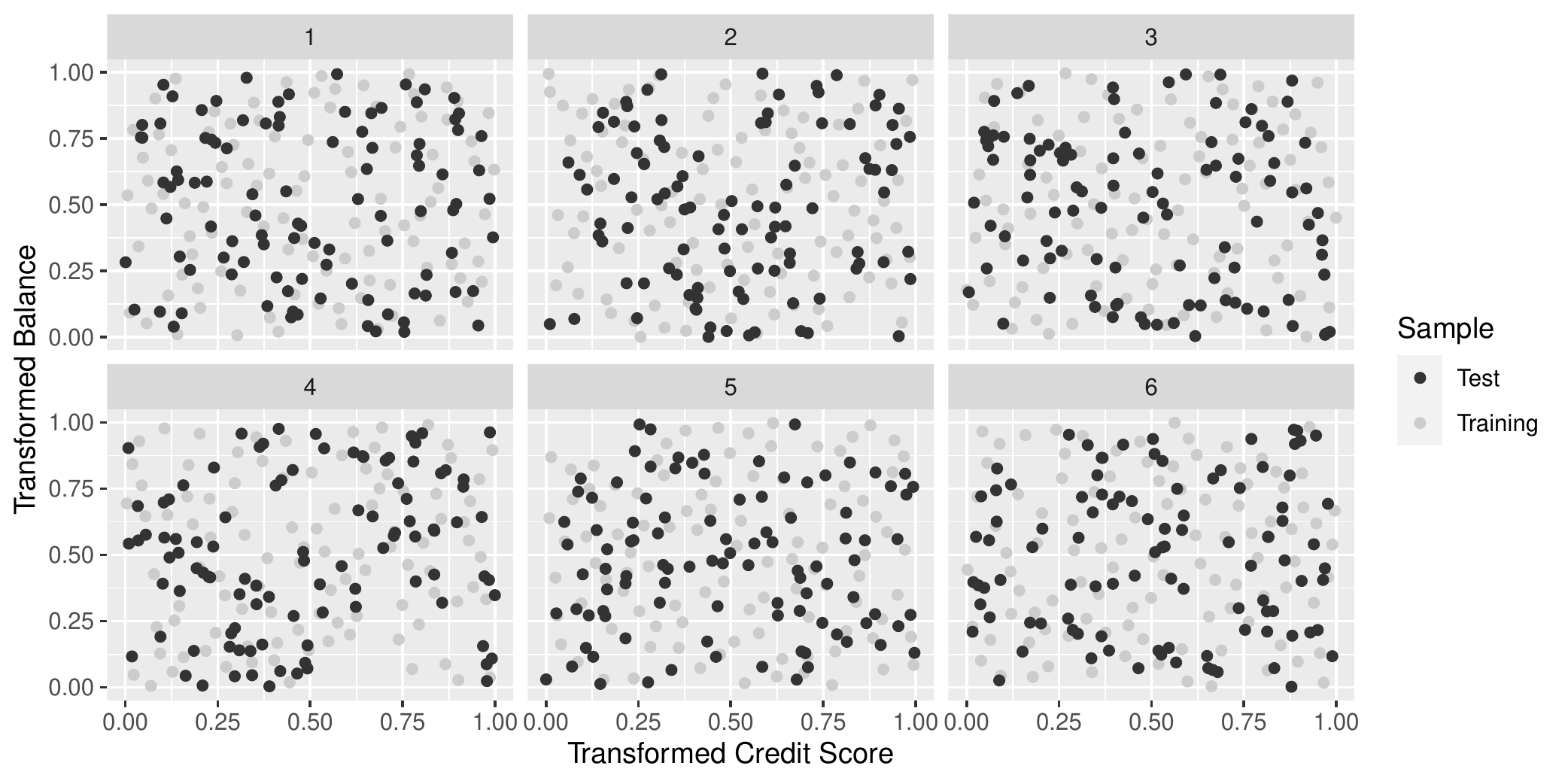}
    \caption{Designs for the training and test sets used in the construction of the GP emulator for the variance of independent accounts. The training design is an optimal sliced Latin hypercube design, with slices defined by the combination of the segment and payment indicator covariates.}
    \label{fig:training-test-design}
\end{figure}

In order to train the emulator we carried out a computer experiment involving extensive simulation from the model. Owing to the mixture of discrete and continuous covariates, the design used for this experiment was an optimal sliced Latin Hypercube design \cite{ba2015optimal}  constructed as follows. First, as is standard in the computer experiments literature, the continuous covariates were transformed to the range $[0,1]$ using the cumulative distribution functions of the distributions specified in Section \ref{sec:detailed-model-description}. This gives transformed variables $\tilde{B} = F_{B}(B)$, $\tilde{C}= F_C(C)$. 
Second, the slices were defined by the six combinations of values of the discrete variables $S\in\{1,2,3\}$ and $Y_0 \in \{0,1\}$, with 100 design points per slice chosen using the R package SLHD \cite{ba2015optimal}, denoted $\mathbf{w}_{syl}= (\tilde{b}_{syl}, \tilde{c}_{syl}, s, y)$, ($l=1,\ldots,100$; $s=1,2,3$; $y=0,1$). 
Finally, as we wish to model the variance of the collections, the variance was estimated for each design point $\mathbf{w}_{syl}$ by taking the sample variance $\hat{v}^2_{syl}$ of \edit{$K$ (we used $K=1000$)} simulated realisations $X^{(k)}_{syl}$ $(k=1,\ldots,K)$ of the total collections for an account with the corresponding covariate values obtained from the inverse transformation, i.e. $(B,C,S,Y_0)= (F^{-1}_B(\tilde{b}_{syl}),F^{-1}_C(\tilde{c}_{syl}), s,y)$. Figure \ref{fig:training-test-design} shows this training design together with an independent test set consisting of 100 random points in each slice.

%
% GP modelling approach - decisions. 
% - independent GP in each slice
% - use log transformation of the response 
% - use unknown constant mean + Matern kernel 
% 
% discussion/justification for decisions:
% - justification for Matern kernel 
% - justification / consequences of log-transformation 
%   - pro: ensures postiive variance predicitons
%   - con: sometimes we get estimated values of zero; these must be excluded from the model fit 
%.   mitigation: these are not 'true' zeroes, but numerical errors corresponding to high kurtosis accounts; it is difficult to accurately quantify the uncertainty in the variance estiamtes for these accounts, so it is entirely natural to exclude them. 

There are many possible approaches for fitting a Gaussian process emulator to these data. \edit{The most important decision that we made in the modelling was to use a log transformation of the variance as the response. The reason for this is that the true account variance is strictly positive, and we wished to preserve this constraint in the predicitons. For the GP prior mean we used an unknown constant and for the covariance function we used the Matérn-5/2 kernel, a common choice for modelling functions that are thought to be at least twice differentiable, see e.g. \cite{stein1999interpolation}, \cite{rasmussen2006gaussian}, \cite{roustant2012dicekriging}. We used the R package DiceKriging \cite{roustant2012dicekriging} to perform the calculations of the GP posterior predictions.}

\edit{The use of the log-transformed variance as the response has some interesting consequences in addition to ensuring positive predictions for the variance function. 
Most pressing, it raises a question of how to treat the reasonably common case of observations for which the observed sample variance is zero, since the corresponding logarithm of negative infinity can not be used in the GP model. Such results typically correspond to an account with an extremely skewed and high kurtosis distribution, in particular with a high probability of zero collections and a very small probability of a large collection whose value is unknown because it was never observed in the simulation. For such accounts the error in the estimate $\hat{v}_{syl}$ is likely large but difficult to quantify and so we argue that such accounts are best excluded from the training set regardless of which transformation is used.} 

\edit{Another consequence of the use of the log transformation is that $\hat{v}_{syl}$ can be regarded as a noisy normal estimate of the true function. In particular Lemma \ref{lemma:dist-logvariance} below, whose proof is given in the appendix, shows that  for large $K$ we can approximate the distribution of the logged variance estimates as 
$$
\log \hat{v}_{syl} 
\sim 
N(
    \log \tilde{v}_{sy}( \tilde{b}_{syl}, \tilde{c}_{syl}), 
    (\kappa_{syl}-1)/K
)\,,
$$
where $\kappa_{syl}$ denotes the kurtosis of the random variable $X^{(k)}_{syl}$ and 
$
\tilde{v}_{sy}(
    \tilde{b},
    \tilde{c}
)
    = 
v(
    F_B^{-1}(\tilde{b}), 
    F_C^{-1}(\tilde{c}),
    s,
    y
)
$ denotes the variance function in the $(s,y)$-slice. This approximation should hold well for accounts that do not suffer from high kurtosis.  The noise variance $(\kappa_{syl}-1)/K$ can be estimated by plugging in the sample kurtosis of the realisations.  }

\begin{lemma}
\edit{The logarithm of the variance estimate $\hat{v}_{syl}$ based on $K$ realisations satisfies
$$
\sqrt{K} \log 
\frac{
    \hat{v}_{syl} 
    }{
    \tilde{v}_{sy}(\tilde{b}_{syl}, \tilde{c}_{syl})
    }
\overset{d}{\to} 
N[0 , 
(\kappa_{syl}-1)]
$$ 
as $K\to \infty$.}
\label{lemma:dist-logvariance}
\end{lemma}

\begin{figure}[htbp]
    \centering
    \includegraphics[width=0.9\linewidth]{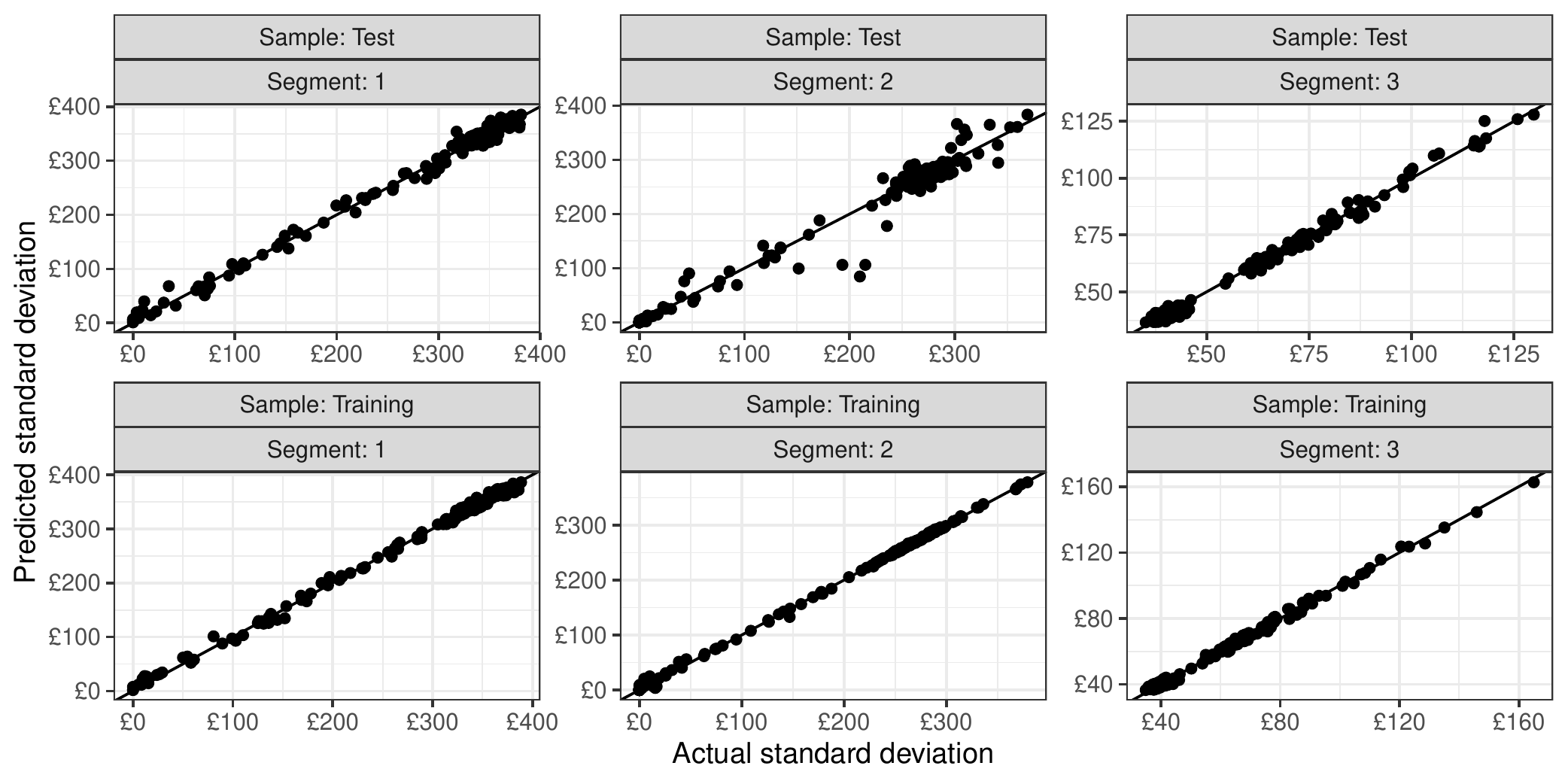}
    \caption{Predicted vs sample standard deviation of the total collections from individual accounts in the training and test sets. Predictions are obtained from the emulation approach in which independent GPs are fitted to the three different segments with log-variance as the response and the following covariates: transformed balance, credit score, and standard deviation of the first month payment indicator.}
    \label{fig:gp-validation}
\end{figure}

\edit{Initially we tried fitting independent Gaussian processes in each $(s,y)$-slice, using transformed balance and credit score as the covariates in each of the slice-specific models.} Although this is appealing in its simplicity, we found that \edit{the} predictions were outperformed by another emulation method. Specifically in this second method, we fit an independent process to each of the three segments (rather than the six slices) with covariates given by the transformed balance, credit score, and also $\sqrt{p_{1}(1-p_{1})}$ where $p_1$ denotes the probability of payment in the first \edit{month}. The latter corresponds to the standard deviation of the payment indicator in the first month. Figure \ref{fig:gp-validation} shows that this alternative method gives effective predictions on the test set. Though a small number \edit{of accounts} in Segment 2 have their variance underpredicted, we believe that the overall quality of the predictions is adequate at the population level for the purposes of uncertainty quantification and informing an  efficient allocation of the computational budget.

\section{Discussion}\label{sec:dicuss}
We have developed uncertainty quantification techniques for simulation-based forecasts of total collections and shown that it is possible to achieve significant reductions in the Monte Carlo error of the estimate of the total expected collections by optimizing the number of simulation realisations according to account-level covariates. Though the estimates of expected total are substantially improved by optimisation of the realisation numbers, the prediction intervals are only marginally smaller due to the fact that the dominant source of uncertainty is the background variability of the realised collections, not the error in estimating expected collections. Alternatively, if one is satisfied that the current levels of Monte Carlo error are acceptable, similar analysis as demonstrated in Section \ref{sec:protect} can be used instead to reduce the computational cost of simulations to achieve the same degree of accuracy, \edit{instead of aiming to further minimise the variance}. Further reductions in computational cost may be achievable through a sampling-based approach in which a representative subset of individuals is considered.
However, from a business perspective there are a multitude of scenarios where a cash value is required for each account, including for audit purposes as well as ad-hoc analyses requiring a re-cut of the segmentations.
Thus we did not consider sampling-based approaches.  

\edit{The methods that we have developed are broadly applicable for computing the expectation of quantities with respect to a large population of heterogeneous stochastic processes, in particular where there are a large number of independent sub-populations which can enable different numbers of realisations to be simulated for each group. The specifics of the paper are focussed on application to modelling the expected collections from portfolios of NPLs, but any model which has the key features of a large heterogeneous population with a large number of independent components would benefit from the methods we present here.}
\section*{Acknowledgements} 
The authors are grateful to Innovate UK and Arrow Global Limited for Knowledge Transfer
Partnership funding (KTP11066), and to Sanjot Gill for chairing the Partnership and for his enthusiastic support from conception to completion of the project. In addition SLC is grateful for a Turing fellowship from the
Alan Turing Institute. 

\section*{Supplementary material}
\editfinal{Supporting R code and numerical results have been made available online at} \url{https://github.com/timwaite/arrow}.

\bibliographystyle{siam}
\bibliography{refs}

\appendix

\section{Proof of Lemma \ref{lemma:dist-logvariance}}
\label{sec:proof-lemma1}
\edit{
\begin{proof}
Standard results, e.g.~\cite{oneill2014some}, p.~294, show that for a variance based on $K$ simulated realisations we have
$$
\sqrt{K}
\left( 
    \frac{  
        \hat{v}_{syl}
        }{
     \tilde{v}_{sy}(\tilde{b}_{syl}, \tilde{c}_{syl})
        } - 1
\right)
\overset{d}{\to} 
N( 0 ,  \kappa_{syl} -1 ) \,
$$
as $K \to \infty$. Applying the delta method, we have that 
$$
\sqrt{K} 
\log 
\frac{
    \hat{v}_{syl}
    }{
    \tilde{v}_{sy}(\tilde{b}_{syl}, \tilde{c}_{syl})
    } 
\overset{d}{\to} 
N(0, \kappa_{syl}-1) \,,
$$
as required.
\end{proof}}

\section{Proof of Theorem \ref{thm1}}
\label{sec:Proof1}

First we present some lemmas establishing asymptotic accuracy of our estimates of the prediction error variance. 

\begin{lemma}
Suppose that regularity conditions (R1)--(R3) hold.
Then 
$
\hat{\sigma}^2_{(X_{\mathcal{I}}-\hat\mu_{\mathcal{I}})}
= 
\sum_{i\in\mathcal{I}} \hat{\sigma}^2_i (1+1/R_i)
$ 
computed with method (M1) as given in Section \ref{sec:UQ} is unbiased for $\Var(X_{\mathcal{I}}-\hat\mu_{\mathcal{I}})$ and has variance of asymptotic order $O(|\mathcal{I}|\bar{\sigma}^4)$.
\label{lemma:order-indep-var}
\end{lemma}

\begin{proof}

Unbiasedness follows from the fact that the $\hat{\sigma}^2_i$ are unbiased for $\sigma^2_i$ ($i\in \mathcal{I}$). As regards the variance, combine the fact that $\Var(\hat{\sigma}^2_i) = \frac{\edit{\sigma^4_i}}{R_i}(\kappa_i - 1 + \frac{2}{R_i-1})$ (e.g. \cite{cho2005variance}) with regularity conditions (R1) and (R2)) to establish that 
\begin{align}
\Var(\hat{\sigma}^2_{(X_{\mathcal{I}}-\hat{\mu}_{\mathcal{I}})}) &= 
		\sum_{i=1}^{N} 
			\left(1+\frac{1}{R_i} \right)^2 
				\frac{\sigma^4_i}{R_i}
				\left(\kappa_i - 1 + \frac{2}{R_i-1}\right) \notag \\
  &\leq 
	   4 (\kappa_\text{max} + 1)  
		 \sum_{i=1}^{N} \sigma^4_i \notag \\
	& \leq
	   |\mathcal{I}| \bar{\sigma}^4 \times 4 (\kappa_\text{max} + 1)
		  (\gamma^2 + 1 ) = O(|\mathcal{I}|\bar{\sigma}^4)\,.
		\label{eq:numerator-ub}
\end{align}
 \end{proof}

\begin{lemma}
Suppose that regularity conditions (R1)--(R3) hold and either 
$
\sigma^2_{\mathcal{D}}(\kappa_{\mathcal{D}}+1) = o(\bar{\sigma}^4|\mathcal{I}|^2)
$ 
or 
$
\sigma^2_{\mathcal{D}}(\kappa_{\mathcal{D}}+1) = O(\bar{\sigma}^4|\mathcal{I}|^2)
$
and $R_{\mathcal{D}}\to \infty$.
Then 
$
\hat{\sigma}^2_{(X-\hat{\mu})}
$
is an unbiased and relatively consistent estimator for $\sigma^2_{(X-\hat{\mu})}$, i.e. 
$
\hat{\sigma}^2_{(X-\hat\mu)}/
\sigma^2_{(X-\hat\mu)}
\to 1
$ 
in probability as $|\mathcal{I}|\to \infty$. 
\label{lemma:relative-consistency}
\end{lemma}

\begin{proof}
Unbiasedness follows from the fact that the estimator is a linear combination of unbiased estimators. Hence the ratio 
$
\rho=
\hat{\sigma}^2_{(X-\hat\mu)}/
\sigma^2_{(X-\hat\mu)}
$
has expectation 1. To establish convergence in probability it is enough to show that $\Var(\rho) \to 0$ as $|\mathcal{I}|\to\infty$. We have
\begin{align*}
    \Var(\rho)&= 
    \frac{
        (1+1/R_{\mathcal{D}})^2 \Var(\hat{\sigma}^2_{\mathcal{D}})
        + \Var(  
            \hat{\sigma}^2_{
              (X_{\mathcal{I}}- \hat{\mu}_{\mathcal{I}})
            }
        )
    }{
         [\sigma^2_{\mathcal{D}} (1+1/R_{\mathcal{D}}) + 
         \sum_{i\in \mathcal{I}} \sigma^2_{i} (1+ 1/R_i) ]^2
    } \\[1ex]
    &\leq 
    \frac{ 4 }{ R_{\mathcal{D}} }
     \frac{
        \sigma^4_{\mathcal{D}}  (\kappa_{\mathcal{D}}+1) 
    }{
        |\mathcal{I}|^2 \bar{\sigma}^4 
    } 
    + \frac{ 
    \Var(  
            \hat{\sigma}^2_{
              (X_{\mathcal{I}}- \hat{\mu}_{\mathcal{I}})
            } 
            )
    }{
       |\mathcal{I}|^2 \bar{\sigma}^4 
    }
\end{align*}
The first term tends to zero due to the conditions in the statement. The second term tends to zero as $|\mathcal{I}|\to \infty$ by Lemma \ref{lemma:order-indep-var}. Hence $\Var(\rho)\to 0$ and the result is proved. 

\end{proof}

\begin{lemma}
Provided that regularity conditions (R1)--(R3) hold
$
\frac{
    X_{\mathcal{I}} - \hat\mu_{\mathcal{I}}
}{
    \sqrt{
      \Var( X_{\mathcal{I}}-\hat\mu_{\mathcal{I}} )
    }
} \to N(0,1)
$
in distribution as $N\to \infty$. 
\label{lemma:normality-indep}
\end{lemma}

\begin{proof}[Proof of Lemma \ref{lemma:normality-indep}]
We first note that the numerator $X_{\mathcal{I}}-\hat\mu_{\mathcal{I}}$ can be written as a sum of independent random variables with expectation zero, namely $X_{\mathcal{I}}-\hat\mu_{\mathcal{I}} = \sum_{i\in \mathcal{I}} \tilde{X}_i$ where $\tilde{X}_i = X_i - \bar{X}_i$ and $\bar{X_i}= \frac{1}{R_i} \sum_{k=1}^{R_i} X_i^k$. The conclusion then follows from the application of the Lyapunov central limit theorem to the sequence $\{\tilde{X}_i\}_{i\in \mathcal{I}}$. To apply the theorem we must verify that the Lyapunov condition holds, i.e. that there exists a $\delta>0$ such that 
\begin{equation}
\lim_{|\mathcal{I}| \to \infty}
	\frac{1}{ \tilde{s}^{2+\delta}_{|\mathcal{I}|}}
	 \sum_{i \in \mathcal{I} } \E|\tilde{X}_{i}|^{2+\delta}
		= 0 \,,
\label{eq:lyapunov}
\end{equation}
where $\tilde{s}^2_{|\mathcal{I}|} = \sum_{i\in \mathcal{I}} \Var(\tilde{X}_i)$. By considering kurtosis, we show that given regularity conditions R1--R3 it suffices to take $\delta=2$.

Recall the behaviour of excess kurtosis under addition or subtraction of independent random variables. Namely, if $Y_1,\ldots,Y_n$ are independent then
$$
\operatorname{Kurt}\left(\sum_{i=1}^{n} Y_i\right) -3 = 
\frac{
	\sum_{i=1}^{n} (\Var Y_i)^2 [\operatorname{Kurt}(Y_i) -3] 
	}{
	 (\sum_{i=1}^{n} \Var Y_i)^{2} \
	} \,.
$$ Thus we have that 
\begin{align*}
\operatorname{Kurt}(\bar{X}_i) -3 &= \frac{\kappa_i -3}{R_i} \\
\operatorname{Kurt}(\tilde{X}_i) -3 &=
 \frac{(\kappa_i - 3)(1+ 1/R_i^3 )}{ (1+1/R_i)^2}  
\end{align*}
and by regularity conditions (R1) and (R2) the kurtosis of each individual term is bounded, say by $\operatorname{Kurt}(\tilde{X}_i) \leq \tilde{\kappa}$, $i\in\mathcal{I}$, for some $\tilde{\kappa}$ not depending on $|\mathcal{I}|$. 
%
% changed up to here  
%
For $\delta=2$, the quantity in \eqref{eq:lyapunov} becomes
\begin{align*}
\frac{
 \sum_{i \in \mathcal{I}} \E( \tilde{X}_i^4)
}{
 ( \sum_{i \in \mathcal{I} } \Var(\tilde{X}_i) )^2
}
 &= 
\frac{
\sum_{i \in \mathcal{I} } \operatorname{Kurt}(\tilde{X}_i) \Var(\tilde{X}_i)^2 }{
 \left( \sum_{i\in \mathcal{I} } \sigma^2_i (1+1/R_i) \right)^2 
} \\
& \leq
\frac{
	\tilde{\kappa} \sum_{i \in \mathcal{I} } \sigma_i^4 (1+ 1/R_i)^2
}{
 (\sum_{ i \in \mathcal{ I} } \sigma^2_i)^2 
}\\
& \leq 
\frac{
  4 \tilde{\kappa}(\gamma^2 + 1)
}{
  \edit{|\mathcal{I}| }
} \,.
\end{align*}
The final upper bound tends to zero as $|\mathcal{I}| \to \infty$ due to regularity condition (R3). Hence the Lyapunov condition holds and the result is proved. 
\end{proof}

\begin{proof}[Proof of Theorem \ref{thm1}]

We aim to show that if the conditions hold then 
$Z= (X-\hat{\mu})/\hat{\sigma}_{(X-\hat{\mu})} \to N(0,1)$ in distribution. Then simple algebra shows that that the interval end-points contain the true total realised collections $X$ with probability equal to 
$$
P\left[ -z_{(1+p)/2} \leq Z \leq z_{(1+p)/2} \right] \to p
$$
as $N\to \infty$. This proves the result.

First suppose that (R1)--(R5a) hold. Then 
\begin{eqnarray}
    Z
    &=& \frac{X-\hat\mu}{\sqrt{\Var(X-\hat\mu)}} 
    \sqrt{\frac{ \Var(X-\hat\mu)}{ \hat{\sigma}^2_{(X-\hat\mu)} } }
    \nonumber \\
     &=& \nonumber
     \frac{(X_{\mathcal{D}} - \hat{\mu}_{\mathcal{D}} )}{
       \sqrt{ \Var(X-\hat\mu) }
     } 
     \frac{1}{\sqrt{\rho}}
     + 
     \frac{(X_{\mathcal{I}} - \hat{\mu}_{\mathcal{I}}) }{
       \sqrt{ \Var(X_{\mathcal{I}}-\hat\mu_{\mathcal{I}}) }
        }
       \sqrt{ 
       \frac{
        \Var(X_{\mathcal{I}}-\hat\mu_{\mathcal{I}})
       }{
        \Var(X-\hat\mu)
       } 
       } 
    \frac{1}{\sqrt{\rho}}\\[1ex]
    &=:& N + Z'\,, 
    \label{eq:NplusZ}
\end{eqnarray}
where we have used the notation $\rho$ from the proof of Lemma \ref{lemma:relative-consistency}.
We can show that $N=o_P(1)$, since
$
E( \sqrt{\rho} N )=0
$ 
and 
$
\Var(\sqrt{\rho} N) = 
\Var(
X_{\mathcal{D}}-\hat{\mu}_{\mathcal{D}}
) / \Var(X-\hat{\mu}) \leq \sigma^2_{\mathcal{D}}(1+1/R_{\mathcal{D}}) / (|\mathcal{I}|\bar{\sigma}^2) \to 0
$, so $\sqrt{\rho} N \overset{p}{\to } 0$ and $N=(\sqrt{\rho} N) \times (1/\sqrt{\rho}) \overset{p}{\to} 0$ by the continuous mapping theorem. Note this argument also shows that  
$
   \frac{
        \Var(X_{\mathcal{I}}-\hat\mu_{\mathcal{I}})
       }{
        \Var(X-\hat\mu)
       } \to 1 
$.
Using this in conjunction with Slutsky's theorem, Lemmas \ref{lemma:relative-consistency} and \ref{lemma:normality-indep}, and the continuous mapping theorem, we note that $Z'\to N(0,1)$ in distribution. A further application of Slutsky's theorem to \eqref{eq:NplusZ} then shows that $Z\to N(0,1)$ as required.

Now suppose that (R1)--(R4) and (R5b) hold. Note that due to condition (R5b)(i)---(iii) we have that
\edit{
\begin{align*}
&\Var(X_{\mathcal{D}}- \hat{\mu}_{\mathcal{D}} )/\sigma^2_{(X-\hat\mu)} \to \beta_1/(\beta_1+\beta_2) = w ; \\
&\Var(\mu_{\mathcal{D}}-\hat{\mu}_{\mathcal{D}})/\sigma^2_{(X-\hat\mu)} \to 0 ; \\
&\Var(X_{\mathcal{I}} - \hat{\mu}_{\mathcal{I}} )/\sigma^2_{(X-\hat\mu)} \to 1-w . \\
\end{align*}}
Now consider  
\edit{
\begin{align*}
\sqrt{\rho} Z
&= \frac{
        X_\mathcal{D}-  {\mu}_\mathcal{D} 
        }{
        \sqrt{
        \Var(X_{\mathcal{D}} - \hat{\mu}_{\mathcal{D}})
        }
        }  \frac{
     \sqrt{
        \Var(X_{\mathcal{D}} - \hat{\mu}_{\mathcal{D}})
        }
   }{
    \sigma_{(X-\hat{\mu}) }
   } 
 \\[1ex]
 &\quad +
   \frac{
      \mu_\mathcal{D}- \hat{\mu}_\mathcal{D} 
        }{
    \sigma_{(X-\hat{\mu}) }
   } 
 \\[1ex]
 &\quad + \frac{ X_{\mathcal{I}} - \hat\mu_\mathcal{I} }{
     \sqrt{\Var(X_{\mathcal{I}} - \hat{\mu}_{\mathcal{I}}) }
    } 
\frac{
     \sqrt{\Var(X_{\mathcal{I}} - \hat{\mu}_{\mathcal{I}}) }
   }{
    \sigma_{(X-\hat{\mu}) }
   }  
 \\[1ex]
 & =: Z_1 + Z_2 + Z_3\,.
\end{align*}}
Above, 
$Z_1$, $Z_2$, and $Z_3$ are mutually independent, with $Z_1 \overset{d}{\to}N(0,w)$ due to (R5b)(iv), $Z_2 \overset{p}{\to} 0$ owing to the variance calculations above, and $Z_3 \overset{d}{\to} N(0,1-w)$ due to Lemma \ref{lemma:normality-indep}. Hence 
$$
(Z_1,Z_2, Z_3)
\overset{d}{\to} 
N\left[ 
    \begin{pmatrix}0\\0\\0\end{pmatrix}\,,
    \begin{pmatrix}
    w & 0 &0 \\ 0&0 &0 \\ 0 & 0 & 1-w
    \end{pmatrix}
\right] \,.
$$
Applying the continuous mapping theorem with $g(z_1,z_2,z_3)= z_1+z_2+z_3$ gives that $\sqrt{\rho}Z \overset{d}{\to} N(0,1)$, and Slutsky's theorem with Lemma \ref{lemma:relative-consistency} gives that $Z \overset{d}{\to} N(0,1)$ as required. 
\end{proof}
%\section{Proof of Theorem \ref{thm2}}\label{sec:Proof2}
%Proofs go here - thanks Tim!

\edit{
\section{Convergence of Algorithm 1}
\label{sec:algo-convergence}

\begin{lemma}
Suppose that the working set $\mathcal{B}$ is such that there are at least two non-active constraints, and  that at least one of the non-active constraints fails to be satisfied under the solution, $\mathbf{R}^\ast(\mathcal{B})$, of the stationarity equations corresponding to $\mathcal{B}$,  in other words $\Var(\hat{\mu}_{j'}) > V_{j'}$ for some $j'\not\in\mathcal{B}$. Then adding $j'$ to the working set strictly increases the variance of $\hat{\mu}_j$ for all other non-actively constrained portfolios.
\label{lemma:algo-alpha-lemma}
\end{lemma} 

\begin{proof}

First note that the solution to the stationarity equations can be written in the form
\begin{align*}
R^\ast_i &= c_i e_j (\mathcal{B}) \,, 
\qquad 
(i \in \mathcal{P}_j)\,,
\\[1ex]
e_j (\mathcal{B}) &= 
\begin{cases}
\epsilon_j \,, & j \in \mathcal{B} \,, \\
\alpha(\mathcal{B}) \,, & j \in \mathcal{B}^c \,,
\end{cases}
\end{align*}
where $c_i >0$ and $\epsilon_j>0$ ($i=1,\ldots,N$; $j=1,\ldots,P$) are independent of the choice of $\mathcal{B}$, specifically by taking
$c_i = \sigma_i$ ($i \in \mathcal{I}$),  $c_i=\sigma_i/\sqrt{|\mathcal{D}_j|}$ $(i \in \mathcal{D}_j$), 
$\epsilon_j =   (\sigma_{\mathcal{D},j} \sqrt{|\mathcal{D}_j|} 
        + \sum_{i \in \mathcal{I}_j} \sigma_i )/V_j$, and 
$\alpha(\mathcal{B}) = C^{\mathcal{B}}_{\text{Rem}} / d^{\mathcal{B}}$.
The proof proceeds by showing that 
\begin{equation}
    \alpha(\mathcal{B}^{+}) < \alpha(\mathcal{B}) \,,
    \label{eq:proof-alpha-inequality}
\end{equation}
where $\mathcal{B}^{+} = \mathcal{B} \cup \{j'\}$ is the augmented working set. 
This enough to establish the result because
\begin{equation}
\Var(\hat{\mu}_{j} ) = 
  \frac{  
        \sigma^2_{\mathcal{D}}  
        }{  
        r_{j} 
        }  
  + \sum_{i \in \mathcal{I}_{j}} 
     \frac{\sigma^2_i}{R_i}
= \frac{\eta_j}{e_j(\mathcal{B})}
=
\begin{cases}
\frac{\eta_j}{\alpha(\mathcal{B})}\,, & j \in \mathcal{B}^c\,, \\[1ex]
\frac{\eta_j}{\epsilon_j} \,, & j \in \mathcal{B}\,, 
\end{cases}
\label{eq:proof-portfolio-var-eqn}
\end{equation}
where $\eta_j>0$ is independent of $\mathcal{B}$, and so for a different non-actively constrained portfolio $j \neq j'$ (i.e. $j\in (\mathcal{B}^{+})^c \subseteq \mathcal{B}^c$) the corresponding variance is $\eta_j/\alpha(\mathcal{B})$ under $\mathbf{R}^\ast(\mathcal{B})$ and $\eta_j/\alpha(\mathcal{B}^{+})$ under $\mathbf{R}^\ast(\mathcal{B}^{+})$, with the latter being strictly larger if \eqref{eq:proof-alpha-inequality} is true.
 
To show \eqref{eq:proof-alpha-inequality} we need some preparatory results, the first of which is that 
\begin{equation}
\epsilon_{j'} > \alpha(\mathcal{B})  \,. 
\label{eq:proof-epsilon-constraint}
\end{equation} 
To see this note that if, as assumed in the conditions of the lemma, the $j'$th constraint is not satisfied then, considering \eqref{eq:proof-portfolio-var-eqn}, we have
$
\frac{ 
    \eta_{j'}
    }{ 
   \alpha(\mathcal{B})
    } 
    > 
    V_{j'}
$. 
Moreover
$
    V_{j'}
 = \frac{
    \eta_{j'}
    }{
    \epsilon_{j'}
    }
$, 
by definition of $\epsilon_{j}$ as the value of $e_j(\mathcal{B})$ when $j$ is included in the active set. Putting the last two statements together and rearranging gives \eqref{eq:proof-epsilon-constraint} as required.

 A second preparatory result is that the change in the remaining budget when we move from working set $\mathcal{B}$ to $\mathcal{B}^{+}$ satisfies
 \begin{equation}
 \epsilon_{j'} \gamma_{j'}=
  C^{\mathcal{B}}_\text{Rem} 
- C^{\mathcal{B}^{+}}_\text{Rem} 
 = \alpha(\mathcal{B}) \sum_{j \in \mathcal{B}^c } \gamma_j 
 - \alpha(\mathcal{B}^{+}) \sum_{j \in (\mathcal{B}^{+})^c} \gamma_j 
 \label{eq:Crem-diff} \,.
\end{equation}
To see this note that the remaining budget can be written in two ways as follows. First, summing over the non-actively constrained portfolios we obtain
\begin{equation}
C^{\mathcal{B}}_{\text{Rem}} 
= 
\sum_{j \in \mathcal{B}^c } \sum_{i \in \mathcal{P}_j }  R_i 
= \sum_{j \in \mathcal{B}^c } \sum_{i \in \mathcal{P}_j }  c_i \alpha(\mathcal{B})
= \alpha(\mathcal{B})  \sum_{j \in \mathcal{B}^c} \gamma_j \,,
\label{eq:Crem-way1}
\end{equation}
where $\gamma_j = \sum_{i \in \mathcal{P}_j} c_i>0$.
On the other hand, subtracting the resource allocated to the actively constrained portfolios from the total budget we obtain
\begin{equation}
C^{\mathcal{B}}_{\text{Rem}} 
= C - \sum_{j\in\mathcal{B}} \epsilon_j \gamma_j \,. 
\label{eq:Crem-way2}
\end{equation}
Taking differences using \eqref{eq:Crem-way1} gives the right hand side of \eqref{eq:Crem-diff}, while use of \eqref{eq:Crem-way2} gives the left hand side.  
 
 Now we have enough information to show that \eqref{eq:proof-alpha-inequality} holds. Eliminating $\epsilon_{j'}$ from  \eqref{eq:Crem-diff} using \eqref{eq:proof-epsilon-constraint} gives
 $$
   \alpha(\mathcal{B}) \gamma_{j'} < \alpha(\mathcal{B}) \sum_{j \in \mathcal{B}^c } \gamma_j 
 - \alpha(\mathcal{B}^{+}) \sum_{j \in (\mathcal{B}^{+})^c} \gamma_j \,,
 $$
 and rearranging we find that 
\begin{equation}
\alpha(\mathcal{B}^{+})
\sum_{j \in (\mathcal{B}^{+})^c} \gamma_j
< 
\alpha(\mathcal{B})
\sum_{j \in (\mathcal{B}^{+})^c } \gamma_j \,,
\label{eq:proof-alpha-inequality-2}
 \end{equation}
 where the right hand side has been obtained using the fact that 
 $
 (\mathcal{B}^{+})^c = (\mathcal{B}\cup \{j'\})^c = \mathcal{B}^c\backslash\{j'\}
 $  and so $\sum_{j \in (\mathcal{B}^{+})^c} \gamma_j = \sum_{j \in \mathcal{B}} \gamma_j - \gamma_{j'}$. Finally, from the conditions in the lemma there were at least two non-active constraints to begin with, so 
 $(\mathcal{B}^{+})^c$ is non-empty. Hence the factor
 $
 \sum_{j \in (\mathcal{B}^{+})^c } \gamma_j
 $ 
appearing in both sides of \eqref{eq:proof-alpha-inequality-2} is positive and so may be cancelled to give the result. 
\end{proof}

\begin{lemma}
Suppose that $\mathcal{B}$ and the associated solution $\mathbf{R}^\ast(\mathcal{B})$ are such that all Lagrange multipliers  corresponding to active constraints are non-negative. Moreover suppose that $\mathcal{B}^+$ is obtained by adding a $j'$ for which the corresponding variance constraint is not satisfied under $\mathbf{R}^\ast(\mathcal{B})$. Then the Lagrange multipliers corresponding to active constraints are also non-negative under $\mathbf{R}^\ast(\mathcal{B}^+)$.
\label{lemma:proof-LM-behaviour}
\end{lemma}

\begin{proof}
First note that 
\begin{equation}
\sqrt{1+\delta^{\mathcal{B}^+}_j} = \epsilon_j/ \alpha(\mathcal{B}^+) > \epsilon_j/\alpha(\mathcal{B})
\label{eq:proof-LM-inequal}
\end{equation}
where the equality is obtained from basic algebraic manipulations of the stationarity equations and the inequality follows directly from Equation \eqref{eq:proof-alpha-inequality} in the proof of Lemma  \ref{lemma:algo-alpha-lemma}. It is therefore enough to show that the right hand side of \eqref{eq:proof-LM-inequal} is at least unity for $j \in \mathcal{B}^+$. For the case $j \in \mathcal{B}$ this follows the fact that the right hand side of \eqref{eq:proof-LM-inequal} equals $\sqrt{1+\delta^{\mathcal{B}}_j}$ combined with the assumption of the lemma that $\delta^{\mathcal{B}}_j \geq 0$. For  $j=j'$ it follows immediately from \eqref{eq:proof-epsilon-constraint}.

\end{proof}

Recall that Slater's condition holds if there is at least one point at which the equality constraints are satisfied and the inequality constraints are strictly satisfied. In our problem this is equivalent to
\begin{equation}
    \sum_{j=1}^{P} \gamma_j \epsilon_j < C \,, 
    \label{eq:slater}
\end{equation}
since $\gamma_j \epsilon_j$ is the minimum amount of the budget that must be spent in the $j$th portfolio to meet the corresponding variance constraint (with equality). 

\begin{lemma}
Suppose that Slater's condition \eqref{eq:slater} holds. If $\mathcal{B}$ contains $P-1$ constraints then the remaining constraint is met under $\mathbf{R}^\ast(\mathcal{B})$.
\label{lemma:proof-slater-lemma}
\end{lemma}

\begin{proof}
Let $j''$ be the single constraint not in $\mathcal{B}$. By Slater's condition \eqref{eq:slater},
$$
C> \sum_{j=1}^{P} \gamma_j \epsilon_j = \sum_{j\in\mathcal{B}} \gamma_j \epsilon_j + \gamma_{j''} \epsilon_{j''} \,.
$$
Combining this with \eqref{eq:Crem-way1} and \eqref{eq:Crem-way2} we see that
$$
\alpha(\mathcal{B}) \gamma_{j''} = C^{\mathcal{B}}_{\text{Rem}} = 
C- \sum_{j\in\mathcal{B}} \gamma_j \epsilon_j > \gamma_{j''} \epsilon_{j''} \,.
$$
Assuming it is positive, the factor $\gamma_{j''}$ above can be cancelled to obtain $\alpha(\mathcal{B})>\epsilon_{j''}$. Combining with \eqref{eq:proof-portfolio-var-eqn} we see that $\Var(\hat{\mu}_{j''}) = \eta_{j''}/\alpha(\mathcal{B}) < \eta_{j''}/\epsilon_{j''} = V_{j''}$, so the variance constraint is satisfied as claimed. 
\end{proof}

\begin{theorem}
If Slater's condition holds then Algorithm 1 converges to a global optimum. 
\end{theorem}
\begin{proof}
First, consider primal feasibility. In each step of the algorithm, we either stop because all primal constraints are satisfied, or we reduce the number of non-active constraints by one. The algorithm must therefore terminate after at most $P-1$ steps, because once we have $P-1$ active constraints the remaining constraint must be met by Lemma \ref{lemma:proof-slater-lemma}. Thus the returned solution will be primally feasible.  

Next consider dual feasibility. At the start of the algorithm there are no active inequality constraints, as $\mathcal{B}=\emptyset$, so the corresponding Lagrange multipliers are zero. Therefore the number of dual constraint violations for the working set starts at zero. Moreover, due to Lemma \ref{lemma:proof-LM-behaviour}, it remains at zero after each iteration. Thus the returned solution will satisfy all dual constraints.

Overall, the solution to which the algorithm converges satisfies both primal and dual feasibility in addition to stationarity and complementary slackness, so it is a KKT point. As the problem is convex, this must be a global optimum.
\end{proof} }

\end{document}